\documentclass[submission, Phys]{SciPost}
\usepackage{bm,amssymb,slashed,graphicx,multirow,soul,mathtools,xspace,array}  
\usepackage{float}   
\allowdisplaybreaks
\usepackage{ bbold }  
\usepackage{ifthen}
\usepackage{subcaption}

\usepackage{algorithm}
\usepackage{algpseudocode}
\usepackage{colortbl}

\usepackage{amssymb}% http://ctan.org/pkg/amssymb
\usepackage{pifont}% http://ctan.org/pkg/pifont

\usepackage{xcolor}
\usepackage{booktabs}
\usepackage{longtable}
\usepackage{cleveref}

\usepackage[labelfont=bf]{caption}

\newcommand{\eg}{\mbox{\itshape e.g.}\xspace}
\newcommand{\ie}{\mbox{\itshape i.e.}\xspace}

\newcommand{\cf}{\mbox{\itshape cf.}\xspace}

\newcommand{\GeV}{\text{GeV}}

\newcommand{\beq}{\begin{equation} }
\newcommand{\eeq}{\end{equation}} 
\newcommand{\bi}{\begin{itemize} }
\newcommand{\ei}{\end{itemize} }
\definecolor{Red}{rgb}{1.,0.,0.}
\definecolor{Grn}{rgb}{0.,0.75,0.}
\definecolor{Blu}{rgb}{0.,0.,1.}
\definecolor{Purp}{rgb}{0.3,0.1,0.6}
\definecolor{Mag}{rgb}{1.,0.,1.}

\newcommand{\cO}{\mathcal{O}}
\newcommand{\cX}{\mathcal{X}}

\newcommand{\package}[2]{%
  \textsc{#1}\ifthenelse{\equal{#2}{}}{\xspace}{~#2\xspace}}
  
\newcommand{\pythia}[1][]{\package{Pythia}{#1}}

\newcommand{\lang}[2]{%
  \texttt{#1}\ifthenelse{\equal{#2}{}}{\xspace}{\texttt{#2}\xspace}}

\DeclareRobustCommand{\Sec}[1]{Sec.~\ref{sec:#1}}

\DeclareRobustCommand{\Tab}[1]{Table~\ref{tab:#1}}

\DeclareMathOperator{\diag}{diag}

\DeclareMathOperator{\Tr}{Tr}

\DeclareRobustCommand{\method}{\textsc{HDSense}\xspace}

\hypersetup{
    colorlinks,
    linkcolor={red!50!black},
    citecolor={blue!50!black},
    urlcolor={blue!80!black}
}

\newcommand{\reportnumber}{FERMILAB-PUB-26-0034-CSAID}
\usepackage[angle=0,scale=1,color=black,firstpage=true,opacity=1]{background}
\SetBgContents{\footnotesize\textsf{\reportnumber}}
\SetBgPosition{current page.north east}
\SetBgHshift{-2in}
\SetBgVshift{-0.5in}
\begin{document}

\begin{center}{\Large 
\textbf{\method: An efficient method for ranking observable sensitivity\\
}}
\end{center}

\begin{center}
Beno\^ it Assi\textsuperscript{1$\spadesuit$},
Christian Bierlich\textsuperscript{2$\clubsuit$},
Rikab Gambhir\textsuperscript{1$\diamond$},
Phil Ilten\textsuperscript{1$\dagger$},
Tony Menzo\textsuperscript{3,4$\star$},
Stephen Mrenna\textsuperscript{1,5$\maltese$},
Manuel Szewc\textsuperscript{1,6$\parallel$},
Michael K. Wilkinson\textsuperscript{1,7$\perp$},
Ahmed Youssef\textsuperscript{1$\ddagger$}, 
and Jure Zupan\textsuperscript{1$\mathsection$}
\end{center}

% Affiliations.
\begin{center}
\textsuperscript{\bf 1} Department of Physics, University of Cincinnati, Cincinnati, Ohio 45221,USA \\
\textsuperscript{\bf 2} Department of Physics, Lund University, Box 118, SE-221 00 Lund, Sweden \\
\textsuperscript{\bf 3} Department of Physics, University of Alabama, Tuscaloosa, AL 35487, USA \\
\textsuperscript{\bf 4} Theory Division, Fermilab, Batavia, Illinois, USA \\
\textsuperscript{\bf 5} Scientific Computing Division, Fermilab, Batavia, Illinois, USA \\
\textbf{\textsuperscript{6}} International Center for Advanced Studies (ICAS), ICIFI and ECyT-UNSAM, 25 de Mayo y Francia, (1650) San Mart\'{i}n, Buenos Aires, Argentina \\
\textsuperscript{\bf 7} Department of Physics,
University of Oklahoma,
Norman, OK 73019, USA\\
${}^\spadesuit${\small \sf assibt@ucmail.uc.edu},
${}^\clubsuit${\small \sf christian.bierlich@hep.lu.se},
${}^\diamond${\small \sf gambhirb@ucmail.uc.edu},
${}^\dagger${\small \sf philten@cern.ch},
${}^\star${\small \sf 
amenzo@ua.edu},
${}^\maltese${\small \sf mrenna@fnal.gov},
${}^\parallel${\small \sf mszewc@unsam.edu.ar},
${}^\perp${\small \sf michael.k.wilkinson@ou.edu},
${}^\ddagger${\small \sf youssead@ucmail.uc.edu},
${}^\mathsection${\small \sf zupanje@ucmail.uc.edu}
\end{center}

% Logo.
\begin{center}
\includegraphics[width=0.4\textwidth]{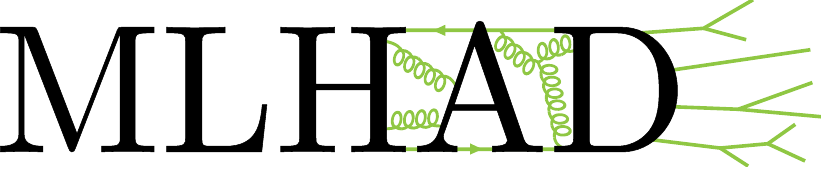}
\end{center}

\begin{center}
\today
\end{center}

\section*{Abstract}
{
Identifying which observables most effectively constrain model parameters can be computationally prohibitive when considering full likelihoods of many correlated observables. This is especially important for, \eg, hadronization models, where high precision is required to interpret the results of collider experiments. We introduce the High-Dimensional Sensitivity (\method) score, a computationally efficient metric for ranking observable sets using only one-dimensional histograms. Derived by profiling over unknown correlations in the Fisher information framework, the score balances total information content against redundancy between observables. We apply \method to rank a set observables in terms of their constraining power with respect to five parameters of the Lund string model of hadronization implemented in \pythia using simulated leptonic collider events at the $Z$ pole. Validation against machine-learning--based full-likelihood approximations demonstrates that \method successfully identifies near-optimal observable subsets. The framework naturally handles data from multiple experiments with different acceptances and incorporates detector effects. While demonstrated on hadronization models, the methodology applies broadly to generic parameter estimation problems where correlations are unknown or difficult to model. 
}

\vspace{10pt}
\noindent\rule{\textwidth}{1pt}
\tableofcontents\thispagestyle{fancy}
\noindent\rule{\textwidth}{1pt}
\vspace{10pt}

\section{Introduction}
\label{sec:intro}

The selection of ``optimal'' observables to constrain model parameters is a fundamental problem in experimental design across sciences.
By the Neyman-Pearson lemma~\cite{10.1098/rsta.1933.0009}, the statistically optimal test statistic for parameter sensitivity is the likelihood function $\mathcal{L}(\theta|\cO)$, where $\cO$ represents the raw observed data or derived observables and $\theta$ the model parameter values.
However, accessing the full likelihood necessitates correct modeling of all systematic uncertainties, as well as the complex correlations between observables, which can be difficult to achieve in practice.
In principle, machine-learning--based methods can approximate this full likelihood~\cite{Cranmer:2015bka,Coccaro:2019lgs,Rizvi:2023mws}, but this often requires expensive simulation and training, large amounts of data, and may be a biased approximation\footnote{See Refs.~\cite{Brehmer:2018kdj, Brehmer:2018eca,Cranmer_sbi_review,deistler2025simulationbasedinferencepracticalguide} for a general discussion in the context of simulation-based inference, and Ref.~\cite{ATLAS:2025clx} for a detailed application of the training techniques with their necessary diagnoses to a realistic measurement.}.
In most practical situations, one therefore must constrain the model parameters with only partial access to the full likelihood,
\eg, the marginal one-dimensional likelihood distributions for each observable, which is a typical example of this approach and one we will focus on here.

As a demonstration, we apply our method to hadronization---the process by which the quarks and gluons (generically ``partons'') produced in high-energy particle collisions transform into the hadrons that are actually observed in detectors.  Because this transition occurs at energy scales where the strong coupling is large, hadronization cannot be calculated perturbatively from first principles and must instead be described by phenomenological models with tunable parameters~\cite{Andersson:1983ia,Andersson:1998tv,Andersson:1997xwk,Field:1982dg, Gottschalk:1983fm, Webber:1983if}.
These hadronization effects enter as systematic uncertainties for a number of precision measurements, such as top quark mass measurements~\cite{ATLAS:2024dxp}, 
determination of the strong coupling constant $\alpha_s$ from event shape observables~\cite{Kluth:2006bw, Bethke:2006ac,ALEPH:2003obs, OPAL:2011aa, Abbate:2010xh, dEnterria:2022hzv, Benitez:2024nav, Hoang:2015hka,Huston:2023ofk}, and $W$ mass measurements at future facilities such as FCC-ee~\cite{deBlas:2025gyz}.  
To control these systematic errors, it is important to have reliable simulations that faithfully reproduce the parton-to-hadron transition with quantifiable uncertainties. 
The two main approaches---the Lund string~\cite{Andersson:1983ia,Andersson:1997xwk} and cluster~\cite{Field:1982dg,Gottschalk:1983fm,Webber:1983if} models---have proven remarkably successful at reproducing experimental data, yet their parameters remain imprecisely determined~\cite{Skands:2014pea,Buckley:2009bj,Buckley:2014ctn}.
This motivates ongoing efforts to better constrain these models, including recent work, by ourselves and by others, using machine learning techniques~\cite{Ilten:2022jfm,Ghosh:2022zdz,Chan:2023ume,Bierlich:2023zzd,Chan:2023icm,Wilkinson:2024jio,Bierlich:2024xzg,Heller:2024onk,Assi:2025avy,Butter:2025wxn, Assi:2025ibi}. 
In this manuscript, we aim to identify observables that are ``most sensitive'' to hadronization model parameters in order to guide decisions about where to invest finite resources in precision measurements in order to maximize the efficacy and efficiency of such machine-learning--based efforts at model constraints.
 
In this paper, we provide an answer to the following question: ``\textit{Given a large set of observables that are  measurable in principle, and knowledge about how each observable is individually affected by model parameters, but not necessarily observable correlations, what is the minimal subset of observables that still gives maximal or close-to-maximal sensitivity to the model parameters?}''
That is, we provide a method for approximating the ``most constraining'' set of observables using only information contained within the marginal distributions of the model.
Our approach is practical: we define a computationally efficient score $\mathcal{S}_{\rm HD}$, the High-Dimensional Sensitivity (\method) score\footnote{``High Dimensional'' because we apply it to $\sim 15$ observables.}, that is used to rank sets of sensitive observables without assuming access to correlation information. Although \method assumes no knowledge of the full likelihood, downstream applications may model it in this reduced subset of observables, with correspondingly reduced computational cost.
 We apply the score to constrain parameters in the Lund string hadronization model (as implemented in the \pythia Monte Carlo event generator~\cite{Bierlich:2022pfr}) given a choice from a number of hadronization-sensitive observables, and we show that the chosen sets of observables constrain the model parameters (nearly) maximally.

An important caveat to our method is that the \method score is \emph{model-dependent}: it quantifies sensitivity assuming a specific model is a good fit to the data. The score uses the Fisher information~\cite{Fisher:1925} to evaluate how well observables can constrain the model parameters in the vicinity of a reference point (typically the best-fit values). 
If the assumed model is incorrect or the true parameters lie far from the reference point, the score may not identify the optimal observable set for discriminating between models or exploring distant regions of parameter space.
In many practical situations, however, where there already exists a model that describes the experimental data reasonably well, this model-dependent approach is appropriate and computationally tractable, and directly applicable to systematic uncertainty estimation. The case of Monte Carlo tuning---that is, calibrating the parameters of a fixed phenomenological model to reproduce a chosen reference dataset using finite generated samples---for a given hadronization model (\eg, string or cluster fragmentation) is one such practical situation.

An additional caveat is that \method is designed to deal with observables that are by themselves sensitive to the model parameters and whose correlation structure is not paramount to their sensitivity. Thus, we assume a certain expert knowledge in the selection of the original, large set of observables. In other words, \method aims to provide a useful selection tool among good observables, not to derive the optimal observables themselves from general representations.

While we demonstrate \method on hadronization model parameters in particle physics, we emphasize that it is in fact applicable more generally to any problem involving parameter estimation from multiple observables with poorly-known or even unknown correlations.
This methodology could be applied, for instance, to optimize observable selection in contexts such as fits of astrophysical models, effective field theory coefficients, parton distribution functions, or other phenomenological model parameters~\cite{Ethier:2021bye,Planck:2018vyg,NNPDF:2021njg,Hou:2019efy,Ellis:2020unq}.

The rest of this paper is organized as follows: 
In \cref{sec:hd_sense}, we define the \method score $\mathcal{S}_{\rm HD}$ as a metric to rank the sensitivity of sets of observables, and we show how to calculate it using binned histograms. 
In \cref{sec:derivation}, we provide an information-theoretic justification for the form of $\mathcal{S}_{\rm HD}$, deriving it from likelihood principles and profiling over unknown correlations, thus contextualizing it as a principled approach to the design of optimal experiments.
In \cref{sec:toys}, we validate the score using toy examples with known correlation structure.
In \cref{sec:application}, we apply \method to rank observables for constraining Lund string model hadronization parameters in \pythia, demonstrating its practical utility and discussing how experimental efficiencies and multiple experiments can be incorporated. 
We conclude in \cref{sec:conclusions} with a summary and outlook.  In appendices we collect the definitions of observables (\cref{sec:observable_definition}), give the results for fixed overlap penalty (\cref{app:fixed_beta}), and discuss the relation to optimal design scores (\cref{app:sec:determinant:score}).
% \jz{Appendices to be added.}

\section{Quantifying sensitivity}
\label{sec:hd_sense}
% %
An observable is ``sensitive'' if it constrains at least some of the parameters of the model, \ie, our measure of sensitivity is based on how the observable responds to changes in the model parameters $\theta_a$, where $a=1,\ldots, N_\text{par}$, with $N_\text{par}$ the number of parameters.
One advantage of such an approach is that this abstracts away the physics, and the selection becomes a purely statistical problem that can be applied to generic models, not only hadronization.

This approach requires the specification of a particular model with explicit parameter dependence. While a model-independent measure would be desirable in principle, it is beyond the scope of our work here; parameter sensitivity is inherently tied to specific model implementations, making our model-dependent approach both necessary and practically useful to our primary goal of selecting the ``best'' observables to invest in measuring precisely.
In our numerical examples, we use the Lund string fragmentation model of hadronization~\cite{Andersson:1983ia,Andersson:1997xwk} as implemented in \pythia~\cite{Bierlich:2022pfr}, though our methodology applies equally well to other hadronization models such as the cluster model~\cite{Field:1982dg,Gottschalk:1983fm,Webber:1983if}, and indeed to any parametric model. 

To select the $K$ ``best'' observables from a total of $N_\text{obs}$, one na\"ively wants, for each subset of $K$ observables $\{\cO_1, \ldots, \cO_K\}$, to compare the likelihoods $\mathcal{L}(\bm \theta|\cO_1, \ldots, \cO_{K})$ and select the set that provides the tightest constraints on $\bm \theta$. 
The challenge is that constructing and numerically exploring such likelihoods is computationally intensive, becoming prohibitively expensive as the number of parameters and observables grows.
We therefore follow an alternative approach: in \cref{sec:score} we construct the {\em \method} score ${\cal S}_\text{HD}$, which can be used to efficiently select the $K$ observables most sensitive to hadronization effects. 
In \cref{sec:single:obs:Fisher} we show how to efficiently calculate the single-observable Fisher information matrices that enter ${\cal S}_\text{HD}$.
\subsection{\method score}
\label{sec:score}

Consider a set of $N_\text{obs}$ observables,  $\bm \cO=\{\cO_1, \ldots, \cO_{N_\text{obs}}\}$, out of which we want to choose $K$ observables that best constrain the parameters $\bm \theta=\{\theta_1,\ldots,\theta_{N_\text{par}}\}$ of the hadronization model. 
If the full likelihood function $\mathcal{L}(\bm \theta|\bm \cO)$ for all possible sets of observables is available, the full Fisher information matrix,
\begin{align}
    I_{ab}(\bm \theta) = \mathbb{E}_{p(\bm \cO | \bm \theta)}\left[\frac{\partial \log p(\bm \cO | \bm \theta)}{\partial \theta_a}\frac{\partial \log p(\bm \cO | \bm \theta)}{\partial \theta_b}\right],
     \label{eq:full_fisher}
\end{align}
where $p(\bm \cO | \bm \theta)$ is the probability density of measuring the observables $\bm \cO$ given the parameters $\bm \theta$, contains all relevant information about parameter constraints~\cite{Fisher:1925}. Intuitively, larger eigenvalues of $I_{ab}$ indicate that the likelihood changes more rapidly with parameter variations, enabling tighter constraints.
Furthermore, the Cram\'er-Rao bound states that the covariance of any unbiased estimator $\hat{\bm \theta}$ of $\bm \theta$ satisfies~\cite{Cramer:1946,Rao:1945}
\begin{align}
\label{eq:cov:CR}
    \text{cov}(\hat{\bm \theta}) \geq I^{-1}(\bm \theta).
\end{align}

To summarize the full Fisher matrix into a single scalar, so that we have a single well-defined optimization problem, we use its trace. (Alternative choices using the determinant are explored in~\cref{app:sec:determinant:score}.) The trace has the advantage of being additive for uncorrelated observables and emphasizes the sum of information across all parameter directions.

In the definition of ${\cal S}_\text{HD}$ we make one crucial simplification. The challenge one often faces is that the full Fisher information matrix \cref{eq:full_fisher} requires knowledge of the complete joint distribution $p(\bm \cO | \bm \theta)$, including all correlations between observables.
In practice, we often only have access to the marginal distributions of each observable, \ie, the one-dimensional probability distributions $p_i(\cO_i | \bm \theta)$ for each observable $\cO_i$ given hadronization parameters $\bm \theta$. The corresponding single-observable Fisher information matrices are
\beq
\label{eq:Iab:def}
     I^{(i)}_{ab}({\bm \theta}) =\mathbb{E}_{p_i(\cO_i | \bm \theta )}\Big[\frac{\partial \log p_i(\cO_i|\bm \theta )}{\partial \theta_a}  \frac{ \partial \log p_i(\cO_i |\bm \theta )}{\partial \theta_b}\Big].
\eeq
Using these single-observable Fisher matrices, we define the \method score 
\beq
    \mathcal{S}_\text{HD}({\mathcal X})
\;=\;
 \;\mathrm{Info}(\mathcal{X})
\;\Big[\,1-\beta\,\mathcal{P}_{\mathrm{overlap}}(\mathcal{X})\,\Big],
\label{eq:bounded-score}
\eeq
where ${\cal X}\subseteq\{\cO_1, \ldots, \cO_{N_\text{obs}}\}$ is any subset of observables, as a proxy for the trace of the full Fisher information matrix.
Here, $\mathrm{Info}(\cal X)$ quantifies the total information content, ${\cal P}_{\rm overlap}(\cal X)$ penalizes redundant constraints, and $\beta$ controls the strength of this penalty. These components are defined below.
In our approach, the best $K$ observables out of $N_\text{obs}$ total observables are those with the largest score ${\cal S}_\text{HD}$.
We will justify the precise form of this ansatz in \cref{sec:derivation}.

\paragraph{Information content.} 
The term $\mathrm{Info}(\cal X)$ measures the total information provided by the observables in $\cal X$:
\begin{align}
    \text{Info}(\cal X) &= \sum_{i \in \cal X} \Tr I^{(i)},
\end{align}
where $I^{(i)}\equiv I^{(i)}( \bm{\hat \theta})$ is evaluated at the maximum likelihood estimate of $\bm \theta$.\footnote{When Fisher information matrices appear without an explicit argument, they are understood to be evaluated at $\bm{\hat \theta}$.}
If there were no correlations between observables, the observables would be independent and $\text{Info}(\cal X)$ would equal the trace of the total Fisher information matrix, $\text{Info}({\cal X})=\Tr I$\footnote{In this work, we consider two observables to be uncorrelated if they are independent, \ie their mutual information is null.}. 
However, observables are generally correlated---in which case combining them may place similar constraints on $\bm \theta$ as using a subset of them. When correlations are known, they can be properly accounted for. In our setting, where correlations are unknown, naively summing single-observable Fisher information matrices overestimates the total information content. We correct for this redundancy via an overlap term.

\paragraph{Overlap penalty.}
To account for redundancy, the ${\cal S}_\text{HD}$ score penalizes overlapping constraints via the term
\begin{equation}
\label{eq:Poverlap}
\mathcal{P}_{\mathrm{overlap}}(\mathcal{X}) 
\;=\;
\frac{2}{\sum_{k\in \mathcal{X}}\mathrm{Tr}[I^{(k)}]}\sum_{\substack{i,j\in \mathcal{X}\\ i<j}}\sqrt{\mathrm{Tr}[I^{(i)}]\mathrm{Tr}[I^{(j)}]\cos(\Phi^F_{ij})},
\end{equation}
where $\cos(\Phi^F_{ij})$ measures the alignment between Fisher information matrices:
\begin{align}
    \cos(\Phi^F_{ij}) &= 
\frac{\langle I^{(i)},I^{(j)}\rangle_F}{||I^{(i)}||_F ||I^{(j)}||_F}\,.
\label{eq:PhiF}
\end{align}
Here, $\langle A, B\rangle_F =\Tr[A^TB]$ is the Frobenius inner product, which measures how aligned matrices $A$ and $B$ are, and $||A||_F^2=\langle A, A\rangle_F$ is the Frobenius norm. 

The form of $\mathcal{P}_{\mathrm{overlap}}(\mathcal{X})$ involves several design choices. In \cref{eq:Poverlap}, we weight each observable pair by $\sqrt{\mathrm{Tr}[I^{(i)}]\mathrm{Tr}[I^{(j)}]}$ to give more weight to overlaps between highly informative observables. We use the trace $\Tr I^{(i)}$ rather than the Frobenius norm $||I^{(i)}||_F = \sqrt{\sum_a \lambda_a^2}$ (where $\lambda_a$ are eigenvalues), because the trace provides a measure of total constraining power across all parameter directions. Both choices emphasize the dominant eigenvalues, the trace as a linear sum $\sum_a \lambda_a$ and the Frobenius norm as a weighted sum favoring larger eigenvalues, but the trace has a more direct interpretation and connects naturally to the profiling derivation in \cref{sec:derivation}. In practice, we find both choices yield similar observable rankings, as Fisher matrices for hadronization observables are typically dominated by a few large eigenvalues.
The overall scale $\frac{2}{\sum_{k\in \mathcal{X}}\mathrm{Tr}[I^{(k)}]}$ arises from scaling the weighted alignment by the total information, making the overlap unitless in the process. It is also further justified by the derivation in \cref{sec:derivation}. One should note however that this pre-factor is effectively absorbed into the $\beta$ definition presented below and thus does not affect the performance.

\paragraph{Selection algorithms.}
The \method score $\mathcal{S}_\text{HD}$ can be used to select the best $K$ observables from $N_\text{obs}$ candidates using two approaches.
For moderate problem sizes, one can directly compute $\mathcal{S}_\text{HD}$ for all ${N_\text{obs} \choose K}$ combinations and select the subset with the highest score. This exhaustive search is feasible when $N_\text{obs}$ is small enough (typically $N_\text{obs} \lesssim 20$), and is the algorithm we use in all studies for observable selection.
For larger problems, we can employ a greedy ``remove-one'' procedure: starting with all $N_\text{obs}$ observables, we iteratively remove the observable whose exclusion least reduces $\mathcal{S}_\text{HD}$, continuing until $K$ observables remain.
This greedy algorithm has computational complexity $\mathcal{O}(N_\text{obs}^2)$ rather than $\mathcal{O}({N_\text{obs} \choose K})$, making it practical for larger observable sets. 
As a byproduct, the order of removal provides a natural ranking of observables by their relative importance.  Thus, we can also use it to order the observables within a given subset of $K$ observables, effectively going from $K$ to $K-1$ greedily. In this work, we only use the greedy algorithm for this ordering, and not for observable selection.

\paragraph{Penalty strength.}
The parameter $\beta$ in \cref{eq:bounded-score} controls how strongly overlaps are penalized, determining the balance between maximizing total information and minimizing redundancy.
As a heuristic choice, we set
\begin{equation}
    \beta = \frac{\beta_0}{\max_{\mathcal{X}}\mathcal{P}_{\mathrm{overlap}}(\mathcal{X})} \quad \text{with~}\beta_0=0.5.
    \label{eq:heuristic}
\end{equation}
Any value $\beta_0\in [0,1]$ could be used, which ensures that the factor $1-\beta\,\mathcal{P}_{\mathrm{overlap}}(\mathcal{X})$ in \cref{eq:bounded-score} remains between 0 and 1. 
The choice $\beta_0=0.5$ represents a compromise between two extremes: $\beta=0$ (no penalty, assuming uncorrelated observables) and $\beta=1$ (maximal penalty for correlations). The former is undesired if we seek to account for correlations and avoid selecting redundant observables, while the latter is also inconvenient since it may avoid selecting observables that are powerful because it over-estimates how redundant they are with other observables in the set. Additionally, we also need to account for how the importance of correlations varies as a function of $K$: for low $K$, since as much constraining power as possible needs to be collected with few observables, the total information should be prioritized over a small overlap between observables. Conversely, the larger $K$ is, improving the variety of the selected subset becomes more important, and thus correlations should be emphasized. 
Because of this, and although in practice the optimal $\beta$ depends on the true but unknown correlation structure, our heuristic provides a reasonable default for the exhaustive algorithm\footnote{The greedy algorithm requires a different heuristic choice to account for the iterative procedure and the reduced set of possible combinations at each $K$.} that we validate empirically in \cref{sec:validation}. 

\subsection{Computing single-observable Fisher information matrices}
\label{sec:single:obs:Fisher}

To compute the \method score $\mathcal{S}_\text{HD}$, we require the single-observable Fisher information matrices $I_{ab}^{(i)}({\bm \theta})$ defined in \cref{eq:Iab:def}.
We compute these matrices numerically by simulating the one-dimensional distribution of observable $\mathcal{O}_i$, which we bin into $B$ bins, so that the $N$ simulated events are distributed into a vector of bin occupancies $\bm{n}=(n_1,\dots,n_{B})$. 

We denote the expectation values for the bin fractions $n_m/N$ for the first $B-1$ bins (keeping $N$ fixed) as $\bm \alpha=(\alpha_1,\dots,\alpha_{B-1})$, so that the expectation value for the fraction of events in the last bin is $\mathbb{E}[n_B]=N\big(1- \sum_{m=1}^{B-1} \alpha_m\big)$. Note that $\bm \alpha$ are functions of the hadronization parameters, $\bm \alpha(\bm \theta)$. We can then compute the Fisher information matrix using the chain rule (dropping the observable index $i$ on the right-hand side to simplify the notation):
\begin{equation}
    \begin{split}
    \label{eq:Iab:chain:rule}
    I_{ab}^{(i)}(\bm \theta)&=\mathbb{E}_{\bm{n}\sim p(\bm{n}|\bm\alpha(\bm \theta))}\Big[\frac{\partial}{\partial\theta_a}  \log p\big(\bm{n}|\bm{\alpha}(\bm\theta)\big)\frac{\partial}{\partial\theta_b}  \log p\big(\bm{n}|\bm{\alpha}(\bm\theta)\big)\Big]\\
&=  \sum_{m,n=1}^{B-1}\frac{\partial\alpha_{m}}{\partial\theta_a} \frac{\partial\alpha_{n}}{\partial\theta_b} \tilde I_{mn}(\bm \alpha),
    \end{split}
 \end{equation}
where the $(B-1)\times (B-1)$ matrix $\tilde I(\bm \alpha)$ is the Fisher information matrix that would arise, if the bin fractions $\alpha_m$ were themselves the parameters of interest,
\beq
\label{eq:tilde:Imn}
\tilde I_{mn}(\bm \alpha)=\mathbb{E}_{\bm{n}\sim p(\bm{n}|\bm \alpha)}\Big[ \frac{\partial}{\partial\alpha_{m}}\log p(\bm{n}|\bm{\alpha})  \frac{\partial}{\partial\alpha_{n}}\log p(\bm{n}|\bm{\alpha}) \Big].
\eeq
The probability of having $\bm n$ events in the $B$ bins, with total number of events $N$ fixed, is given by the multinomial distribution
\begin{equation}
    p(\bm{n}|\bm{\alpha})= \Big(1-\sum_{p=1}^{B-1}\alpha_{p}\Big)^{n_{B}}\prod_{m=1}^{B-1}\alpha^{n_m}_m .
\end{equation}
Note that $n_B$ is not independent, but rather given by $n_B=N-\sum_{m=1}^{B-1}n_m$. 

This probability distribution governs the expectation values in \cref{eq:Iab:chain:rule,eq:tilde:Imn}. Evaluating \cref{eq:tilde:Imn} explicitly, we find
\begin{equation}
    \begin{split}
    \label{eq:tildeImn}
   \tilde I_{mn}(\bm{\alpha})&=\mathbb{E}_{\bm{n}\sim p(\bm{n}|\bm{\alpha})}\Big[\Big(\frac{n_m}{\alpha_m}-\frac{n_{B}}{1-\sum\alpha_p}\Big)\Big(\frac{n_n}{\alpha_n}-\frac{n_{B}}{1-\sum\alpha_p}\Big)\Big]\\
 &=N\Big(\delta_{mn}\frac{1}{\alpha_m}+\frac{1}{1-\sum\alpha_p}\Big),
    \end{split}
\end{equation}
where in the derivation we used the multinomial expectation values
\begin{align}
\mathbb{E}_{\bm{n}\sim p(\bm{n}|\bm{\alpha})}[n_m n_n]&=(N^2 -N) \alpha_m \alpha_n +N \alpha_m \delta_{mn},
\\
\mathbb{E}_{\bm{n}\sim p(\bm{n}|\bm{\alpha})}[n_m n_B]&=(N^2 - N)\alpha_m \Big(1-\sum_{p=1}^{B-1}\alpha_p\Big) , \quad \text{and}
\\
\mathbb{E}_{\bm{n}\sim p(\bm{n}|\bm{\alpha})}[n_B n_B]&=(N^2 -N) \Big(1-\sum_{p=1}^{B-1}\alpha_p\Big)^2 +N \Big(1-\sum_{p=1}^{B-1}\alpha_p\Big).
\end{align}
Note that since $ \tilde I_{mn}(\bm{\alpha})$ is effectively a variance, the $(N^2-N)$ terms cancel in the final expression in \eqref{eq:tildeImn}, leaving us
with the expected $\mathcal{O}(N)$ scaling for a Fisher information matrix based on $N$ observations.

\paragraph{Numerical implementation.}
In \cref{sec:application}, where we apply the score to the problem of hadronization, to compute $\bm{\alpha}(\bm\theta)$ and the gradients  ${\partial\alpha_m}/{\partial \theta_a}$, we use {\pythia} Monte Carlo simulations. For the bin fractions, we use the maximum likelihood estimator based on the number of events that fall in bin $m$ for observable $\cO_i$,
\beq
\alpha_m \simeq \frac{n_m}{N}.
\eeq
For the gradients ${\partial\alpha_m}/{\partial \theta_a}$ entering \cref{eq:Iab:chain:rule}, we exploit the fast event reweighting capability implemented in {\pythia}~\cite{Bierlich:2023fmh,Assi:2025gog}. We fit a linear model
\begin{equation}
    \alpha_{m}(\bm \theta')\simeq \alpha_{m}(\bm \theta)+\sum_{a}(\theta'_a-\theta_{a})\frac{\partial\alpha_m}{\partial \theta_a},
\end{equation}
to the weighted maximum-likelihood estimators
\begin{equation}
    \alpha_{m}(\bm\theta')\simeq \frac{n_{m}(\bm\theta')}{N} =\frac{\sum_{k=1}^{N}w_{k}(\bm \theta', \bm \theta)\Theta(\mathcal{O}_{ik}\in \text{bin}~m)}{\sum_{n=1}^{N}w_{k}(\bm \theta', \bm \theta)}.
    \label{eq:alpha_m:theta'}
\end{equation}
Here, $\Theta$ is the indicator function equal to 1 if the argument is true and 0 otherwise, the sum runs over events indexed by $k$ with the observable $\mathcal{O}_{i}$ taking the value $\mathcal{O}_{ik}$, and $w_{k}(\bm \theta', \bm \theta)$ is the reweighting factor for event $k$ when varying parameters from $\bm \theta$ to $\bm \theta'$ (by definition, $w_{k}(\bm \theta, \bm \theta)=1$). The parameter variations $\bm\theta'$ are chosen close enough to the nominal $\bm\theta$ that the reweighting is accurate and the linear approximation holds. By fitting to multiple reweighted histograms, we extract the gradients needed for \cref{eq:Iab:chain:rule}.

%%%%%%%%%%%%%%%%%%%%%%%%%%%%%%%%%%%%%%%%%%%%%%%%
\section{Theoretical foundation of the \method score}
\label{sec:derivation}

Before presenting numerical examples, we provide a theoretical justification for the \method score defined in \cref{eq:bounded-score}.
We will show that $\mathcal{S}_\text{HD}$ approximates the trace of an ``effective Fisher information matrix,'' obtained by profiling over the unknown correlation structure between observables.
Our strategy is to start with the full Fisher information matrix, which in principle contains complete information about parameter constraints, and then systematically account for our ignorance about correlations.

Throughout this derivation, we will make a series of approximations and work in idealized Gaussian limits. 
These assumptions may not hold exactly for realistic distributions, in which case the \method score may not be fully optimal. 
Nevertheless, the derivation provides insight into the structure of $\mathcal{S}_\text{HD}$ and justifies the specific functional form of the overlap penalty.

\subsection{Setup}

Consider the full negative log-likelihood for all $N_\text{obs}$ observables as a function of the model parameters $\bm\theta$.\footnote{In this work, we are primarily interesting in the parameters of the Lund string model, but we keep our discussion in this section to arbitrary models with parameters $\bm\theta$ and $N_\text{obs}$ observables $\mathcal{O}_{n}$.} Near its minimum at the maximum likelihood estimate $\bm{\hat \theta}$, this has the quadratic form
\begin{align}
    \label{eq:quadratic-expansion}
    -2\mathcal{L}(\bm\theta | \mathcal{O}_1, \mathcal{O}_2, \ldots, \cO_{N_\text{obs}}) \approx [\text{constant}] +  (\theta-\hat{\theta})_a I_{ab}(\theta-\hat{\theta})_b,
\end{align}
where $I_{ab}$ is the full Fisher information matrix for all observables. This matrix determines the covariance of parameter estimates through the Cram\'er-Rao bound, \cref{eq:cov:CR}, and thus quantifies how well $\bm \theta$ can be constrained.\footnote{The use of the quadratic expansion in \cref{eq:quadratic-expansion} is one reason the \method method only works in the region near the reference value of the model.}

In \cref{sec:score} we considered selecting a subset of $K$ observables ${\mathcal X}=\{\cO_1, \ldots, \cO_K\}$ from the full set. The information this subset provides about $\bm\theta$ is encoded in the joint probability distribution $p({\mathcal X}|\bm \theta) \equiv p(\cO_1, \ldots, \cO_K | \bm \theta)$. 

By Sklar's theorem~\cite{Sklar1959}, any joint distribution can be decomposed as
\begin{align}
    p(\cO_1, \ldots, \cO_K | \bm \theta) = c_{\bm \eta}(F_1(\cO_1|\bm\theta), \ldots, F_K(\cO_K|\bm\theta))\, \prod_{i = 1}^K p_i(\cO_i | \bm \theta), 
    \label{eq:copula:decomp}
\end{align}
where $p_i(\cO_i | \bm \theta)$ are the single-observable marginal distributions, $F_i(\cO_i|\bm\theta)$ are the corresponding cumulative distribution functions, and $c_{\bm \eta}$ is the \emph{copula density} that encodes all dependence structure between the observables. (For notational simplicity, in what follows we will write $c_{\bm \eta}(\cX; \bm \theta)$ to denote the copula density evaluated at the appropriate CDFs, suppressing the explicit dependence on $F_i$.)

Note that for a fixed set of observables and parameters, the copula is uniquely determined. However, \emph{we do not know this copula}---our only access is to the one-dimensional marginal distributions $p_i(\cO_i | \bm \theta)$ from individual histograms. To proceed, we parametrize the correlation structure by nuisance parameters $\bm \eta=\{\eta_1, \ldots, \eta_{N_\text{cor}}\}$ that control the dependence between observables. Profiling over these nuisance parameters will allow us to derive a score that depends only on the known marginals. The profiling procedure provides a principled way to understand how correlations affect observable selection, even though the actual correlation structure remains unknown.

\subsection{Profiling over correlations}

To marginalize over the unknown correlation structure, we enlarge the parameter space from just the model parameters $\bm \theta$ to include the nuisance parameters $\bm \eta$ that encode correlations. The full Fisher information matrix for the subset of $K$ observables ${\cal X}$ then becomes a $(N_\text{par}+N_\text{cor})\times (N_\text{par}+N_\text{cor})$ matrix with block structure
\begin{align}
    I_{\rm full} &= \begin{pmatrix}
        I_{ \theta  \theta} & I_{\theta \eta} \\
        I_{ \eta  \theta} &  I_{ \eta  \eta}
    \end{pmatrix}.
\end{align}
Standard profile likelihood methods give the profiled Fisher information matrix for $\bm\theta$ alone:
\begin{align}
    I_{\rm pr.}(\bm \theta) = I_{ \theta  \theta}(\bm \theta, \bm{\hat \eta}) - I_{ \theta \eta}(\bm \theta, \bm{\hat \eta}) I_{ \eta  \eta}^{-1}(\bm \theta, \bm{\hat \eta}) I_{\eta \theta}(\bm \theta, \bm{\hat \eta}), \label{eq:profiled_fisher}
\end{align}
where $\bm{\hat \eta}$ is the maximum likelihood estimate of the correlation parameters for given $\bm\theta$. We define the profiled sensitivity score:\footnote{This score is related to, but not the same as, the $A$ optimality criterion in optimal experimental design literature~\cite{Chaloner:1995,Huan2024OED}, which we discuss further in \cref{app:sec:determinant:score}.}
\begin{align}
    \mathcal{S}_\text{pr.} &= \Tr[I_\text{pr.}].
    \label{eq:S:pr}
\end{align}
In principle, this profiled score contains all information about parameter constraints that can be extracted from the marginal distributions alone. However, computing it requires specifying a parametric form for the copula and evaluating $\bm{\hat\eta}$, which is impractical.
To proceed, we make two assumptions. First, we assume that the observables $\cO_i$ are normally distributed scalars with means $\mu_i(\bm\theta)$ and covariance matrix $\Sigma_{ij}$.
This Gaussian approximation is justified by the central limit theorem: with a large number $N$ of independent measurements, the distribution of any summary statistic (including binned histograms) approaches a multivariate Gaussian. In particular, we can work with scalar summary statistics of each observable distribution, such as the mean or the median over all events in a dataset.\footnote{In Section~\ref{sec:score}, we work with binned observables where the bin counts follow a multinomial distribution. For large event counts, this is well-approximated by a multivariate Gaussian. In this derivation, we work with scalar summary statistics that can be derived from these multivariate Gaussians.} The Gaussian assumption implies that the copula density $c_{\bm \eta}$ encodes only pairwise correlations (no higher-order dependence).
Second, we assume that the covariance matrix $\Sigma_{ij}$ is independent of the model parameters $\bm\theta$. This means that the variations in $\bm\theta$ shift the means $\mu_i(\bm\theta)$ but do not change the widths or correlations. Under these assumptions, the Fisher information simplifies considerably.
The $I_{\theta \theta}$ block in \cref{eq:profiled_fisher} becomes
\begin{align}
   \big[ I_{\theta \theta}(\bm \theta, \bm{\hat \eta})\big]_{ab} &= \partial_{\theta_a} \mu_i(\bm \theta) \, \Sigma^{-1}_{ij} \, \partial_{\theta_b} \mu_j(\bm \theta),
\end{align}
while the cross terms $I_{\theta\eta}$ and $I_{\eta\theta}$ vanish,  $I_{\eta \theta}  = I_{\theta\eta}=0$. This follows because derivatives of the log-likelihood with respect to correlation parameters involve odd central moments of the observables, which are zero for a Gaussian distribution. Thus, the second term in \cref{eq:profiled_fisher} drops out.
Writing the covariance matrix in a factorized form,
\begin{align}
    \Sigma(\bm \eta) = V \rho(\bm \eta) V,
\end{align}
where $V = \diag(\sigma_1, \ldots,\sigma_K)$ contains the standard deviations and $\rho$ is the correlation matrix, the profiled score in \cref{eq:S:pr} becomes
\begin{align}
    \mathcal{S}_{\rm pr.} = \sum_i  \frac{|\partial_{\bm \theta} \mu_i(\bm \theta)|^2}{\sigma_i^2}(\rho^{-1})_{i i} + 2\sum_{i < j}\frac{(\partial_{\bm \theta} \mu_i(\theta)) \cdot (\partial_{\bm \theta} \mu_j(\theta))}{\sigma_i \sigma_j}(\rho^{-1})_{i j}.
    \label{eq:Spr:initial}
\end{align}
Here, vector norms and dot products are taken over the parameter space $\bm \theta$.
Note that the first term contains the traces of single-observable Fisher information matrices. For a Gaussian observable with mean $\mu_i(\bm\theta)$ and variance $\sigma_i^2$, the Fisher matrix is
\beq
\label{eq:simplified_fisher}
[I^{(i)}]_{ab}=\frac{1}{\sigma_i^2} (\partial_{\theta_a} \mu_i )(\partial_{\theta_b} \mu_i), \text{~and thus~} \Tr I^{(i)}=\frac{|\partial_{\bm \theta} \mu_i(\bm\theta)|^2}{\sigma_i^2}.
\eeq
For the second term in \cref{eq:Spr:initial}, we note that
\beq
[I^{(i)T}I^{(j)}]_{ab}=\frac{ (\partial_{\bm \theta} \mu_i )\cdot (\partial_{\bm \theta} \mu_j)}{\sigma_i^2\sigma_j^2}(\partial_{\theta_a} \mu_i )(\partial_{\theta_b} \mu_j), \quad \text{giving} \quad \Tr[I^{(i)T}I^{(j)}] =\frac{ \left[(\partial_{\bm \theta} \mu_i )\cdot (\partial_{\bm \theta} \mu_j)\right]^{2}}{\sigma_i^2\sigma_j^2}.
\eeq
We can express this in terms of the individual traces and an angle $\Phi_{ij}$ via
\beq
 \Tr[I^{(i)T} I^{(j)}] = \Tr[I^{(i)}]\Tr[I^{(j)}] \cos^2(\Phi_{ij}).
\eeq
In the special case where each observable is a scalar (single summary statistic) and under the Gaussian approximation with no parameter-dependent covariance, the Fisher matrices $I^{(i)}$ as defined in \cref{eq:simplified_fisher} are rank-1 (because they each are an outer product of a vector with itself), and their Frobenius norms equal their traces: $||I^{(i)}||_F =\Tr I^{(i)}$. More generally, we can define $\xi_i = \Tr[I^{(i)}]/||I^{(i)}||_F$ to quantify the effective dimensionality. The angle $\Phi_{ij}$ is then related to the Frobenius overlap angle $\Phi_{ij}^F$ defined in \cref{eq:PhiF} by
\beq
 \cos(\Phi_{ij})=\pm \sqrt{\frac{\cos(\Phi_{ij}^F)}{\xi_i \xi_j }},\label{eq:Phi:relation}
\eeq
where the sign is determined by the sign of the scalar product in \cref{eq:Spr:initial}. We are primarily interested in the rank-1 ($\xi_i = 1$) case due to the Gaussian limit.

Using these relations, we can rewrite the profiled score for the rank-1 case, $\xi_i = 1$ for all $i$, as
\begin{equation}
\begin{split}
     \mathcal{S}_{\rm pr.}({\cal X}) =& \left(\sum_i  \Tr[I^{(i)}](\rho^{-1})_{i i}\right)\left(1 + 2\frac{\sum_{i < j}\sqrt{\Tr[I^{(i)}]\Tr[I^{(j)}]}\cos(\Phi_{ij})[\rho^{-1}]_{ij}}{\sum_i  \Tr[I^{(i)}](\rho^{-1})_{i i}}\right)\\
     =& \sum_i  \Tr[I^{(i)}](\rho^{-1})_{i i} + 2 \sum_{i < j}\sqrt{\Tr[I^{(i)}]\Tr[I^{(j)}]}\cos(\Phi_{ij})[\rho^{-1}]_{ij}. \label{eq:score_with_rho}
\end{split}
\end{equation}
This expresses the profiled score entirely in terms of the single-observable Fisher matrices $I^{(i)}$ and the unknown correlation matrix $\rho^{-1}$. The \method score ${\cal S}_\text{HD}$ defined in \cref{eq:bounded-score} will emerge by making a final approximation to handle our ignorance of $\rho$.
%%%%%%%%%%%%%%%%%%%%%%%%%%%%%%%%
\subsection{Parameterizing our ignorance}

To evaluate \cref{eq:score_with_rho}, we need the inverse correlation matrix $\rho^{-1}$, which requires information beyond the single-observable marginal distributions. 
However, we can derive an approximate bound on $\mathcal{S}_{\rm pr.}({\cal X})$ and $\mathcal{S}_{\rm HD}({\cal X})$ by absorbing the unknown correlation structure into a single hyperparameter $\beta$.

First, let us bound the off-diagonal contributions $(\rho^{-1})_{ij}$.
For any positive-definite correlation matrix, the partial correlation inequality gives
\begin{align}
    |(\rho^{-1})_{ij}| \leq \sqrt{(\rho^{-1})_{ii}(\rho^{-1})_{jj}},
\end{align}
which implies
\begin{align}
    \cos(\Phi_{ij})(\rho^{-1})_{ij} \geq -|\cos(\Phi_{ij})| \sqrt{(\rho^{-1})_{ii}(\rho^{-1})_{jj}}= -\sqrt{\cos(\Phi^F_{ij})} \sqrt{(\rho^{-1})_{ii}(\rho^{-1})_{jj}},\nonumber
\end{align}
where in the last equality we have used $|\cos(\Phi_{ij})| = \sqrt{\cos(\Phi^F_{ij})}$ (see \cref{eq:Phi:relation} for $\xi_i = 1$ due to the Gaussian approximation). 
Substituting into \cref{eq:score_with_rho} yields
\begin{equation}
     \mathcal{S}_{\rm pr.} \geq \sum_i  \Tr[I^{(i)}](\rho^{-1})_{i i} - 2\sum_{i < j}\sqrt{\Tr[I^{(i)}]\Tr[I^{(j)}]\cos(\Phi^F_{ij})(\rho^{-1})_{ii}(\rho^{-1})_{jj}}.
     \label{eq:beta_prime}
\end{equation}
This inequality is rigorous but still depends on the unknown diagonal elements $(\rho^{-1})_{ii}$.

Next, we use that $(\rho^{-1})_{ii} \geq 1$ for all $i$ implies\footnote{(Single-observable) Fisher information matrices are positive semi-definite and so have traces greater than or equal to zero.}
\beq
\sum_i  \Tr[I^{(i)}](\rho^{-1})_{i i}\geq \sum_i  \Tr[I^{(i)}].
\label{eq:trineq}
\eeq
We can place an upper-bound on the second summation in \cref{eq:beta_prime} by identifying  the maximum value of $\sqrt{(\rho^{-1})_{ii}(\rho^{-1})_{jj}}$:
\begin{align}
\sum_{i<j}\sqrt{\Tr[I^{(i)}]\Tr[I^{(j)}]\cos(\Phi^F_{ij})(\rho^{-1})_{ii}(\rho^{-1})_{jj}} \leq M \sum_{i<j}\sqrt{\Tr[I^{(i)}]\Tr[I^{(j)}]\cos(\Phi^F_{ij})},
\end{align}
where $M \equiv \max_{i,j}\sqrt{(\rho^{-1})_{ii}(\rho^{-1})_{jj}}$, so that~\cref{eq:beta_prime} gives
\begin{equation}
\mathcal{S}_{\rm pr.} \geq \sum_i \Tr[I^{(i)}](\rho^{-1})_{ii} - 2M\sum_{i<j}\sqrt{\Tr[I^{(i)}]\Tr[I^{(j)}]\cos(\Phi^F_{ij})}.
\label{eq:intermediate_bound}
\end{equation}

To proceed to a computable form, we now assume
\begin{align}
\sum_i \Tr[I^{(i)}](\rho^{-1})_{ii} \geq 2M\sum_{i<j}\sqrt{\Tr[I^{(i)}]\Tr[I^{(j)}]\cos(\Phi^F_{ij})}.
\label{eq:key_assumption}
\end{align}
This assumption holds when correlations are moderate (so that $(\rho^{-1})_{ii}$ and $M$ do not become too large) or when individual observables carry sufficient information relative to their overlaps. Under this assumption, we can take $\Tr[I^{(i)}](\rho^{-1})_{ii}$ as a common factor on the sum, replace it with its lower bound $\sum_i \Tr[I^{(i)}]$ and define the effective hyperparameter
\beq
\beta \equiv \frac{\sum_i  \Tr[I^{(i)}]}{\sum_i  \Tr[I^{(i)}] (\rho^{-1})_{ii}} M,
\label{eq:beta_def}
\eeq
yielding
\beq
\label{eq:Spr:SHD:ineq}
 \mathcal{S}_{\rm pr.} \gtrsim \left(\sum_i  \Tr[I^{(i)}]\right)\left(1 - 2\beta\frac{\sum_{i < j}\sqrt{\Tr[I^{(i)}]\Tr[I^{(j)}]\cos(\Phi^F_{ij})}}{\sum_i  \Tr[I^{(i)}]}\right) \equiv {\cal S}_\text{HD}. 
 %\label{eq:inequality}
\eeq

Thus, the profiled Fisher information score $\mathcal{S}_{\rm pr.}$ defined in \cref{eq:S:pr} is approximately bounded from below by the \method score ${\cal S}_\text{HD}$ proposed in \cref{eq:bounded-score}. Our ignorance about the correlation structure has been absorbed into the hyperparameter $\beta$.

Moreover, even if \cref{eq:key_assumption} does not hold, we can apply \cref{eq:trineq} directly to obtain the more conservative bound
\begin{equation}
\begin{split}
\mathcal{S}_{\rm pr.} \geq& \sum_i  \Tr[I^{(i)}] - 2 M \sum_{i<j}\sqrt{\Tr[I^{(i)}]\Tr[I^{(j)}]\cos(\Phi^F_{ij})} \\
=& \left(\sum_i  \Tr[I^{(i)}]\right)\left(1 - 2M\frac{\sum_{i < j}\sqrt{\Tr[I^{(i)}]\Tr[I^{(j)}]\cos(\Phi^F_{ij})}}{\sum_i  \Tr[I^{(i)}]}\right) \approx {\cal S}_\text{HD},
\end{split}
\end{equation}
which has a similar functional form to ${\cal S}_\text{HD}$, with the unknown correlation structure now absorbed entirely into $M$. Thus, in either case, the profiled score $\mathcal{S}_{\rm pr.}$ is approximately bounded from below by ${\cal S}_\text{HD}$, with $\beta$ playing the role of $M$ (up to normalization factors). Since we do not know $M$ in practice, we treat $\beta$ as a tunable hyperparameter using the heuristic in \cref{eq:heuristic}.

Note that $\beta$ need not be less than 1, as the factor $\frac{\sum_i \Tr[I^{(i)}]}{\sum_i \Tr[I^{(i)}](\rho^{-1})_{ii}} \leq 1$ while $M$ can be arbitrarily large for strong correlations. The only constraint on $\beta$ is that the second factor in \cref{eq:Spr:SHD:ineq} must remain positive for ${\cal S}_\text{HD}$ to represent a meaningful score; the relationship to $\mathcal{S}_{\rm pr.}$ would become meaningless otherwise. In practice, since we do not know the true $(\rho^{-1})_{ii}$ values, we use the heuristic choice $\beta = 0.5/\max_{\mathcal{X}}\mathcal{P}_{\mathrm{overlap}}(\mathcal{X})$ defined in \cref{eq:heuristic}, which we validate empirically in \cref{sec:validation}.

In deriving this score, two key assumptions were made: that the underlying distribution is Gaussian, and that its correlation structure (though not necessarily its variance) is parameter-independent. This is only necessary for motivating the score, and it is in this limit where the score is expected to produce the tightest information bound and the most optimal observable selections. However, we emphasize that neither of these conditions are required to use \method in realistic settings, even if the resulting score may become suboptimal.

%%%%%%%%%%%%%%%%%%%%%%%%%%%%%%%%%%%%%%%%%%
\section{Toy study: perfectly correlated Gaussians}\label{sec:toys}

As a test and to illustrate how the \method score works, we apply the \method score in an example where correlations are fully known.
As the simplest nontrivial example we consider a set of independent observables, copied several times, inducing perfect correlations between the copies, thus resulting in a \emph{block-diagonal} likelihood.
\method should be able to select mutually independent observables out of the full set of observables that also contains copies.

The observables $\mathcal{O}_i$ are assumed to form a multivariate Gaussian,
\begin{align}
    p(\mathcal{O}_1, ... \mathcal{O}_K) &\propto e^{-\frac{1}{2}(\mathcal{O}-\mu(\bm \theta))_i \Sigma_{ij}^{-1} (\mathcal{O}-\mu(\bm \theta))_j}, \label{eq:multivar_gaussian}
\end{align}
where $\mu_i(\bm \theta)$ are parameter-dependent means and $\Sigma$ is the covariance matrix.
Without loss of generality, the variances of the individual observables are assumed to be $\sigma^2 = 1$, which can be achieved by appropriately rescaling the observables, if needed.

\begin{figure}[t]
\centering
\includegraphics[width=0.41\textwidth]{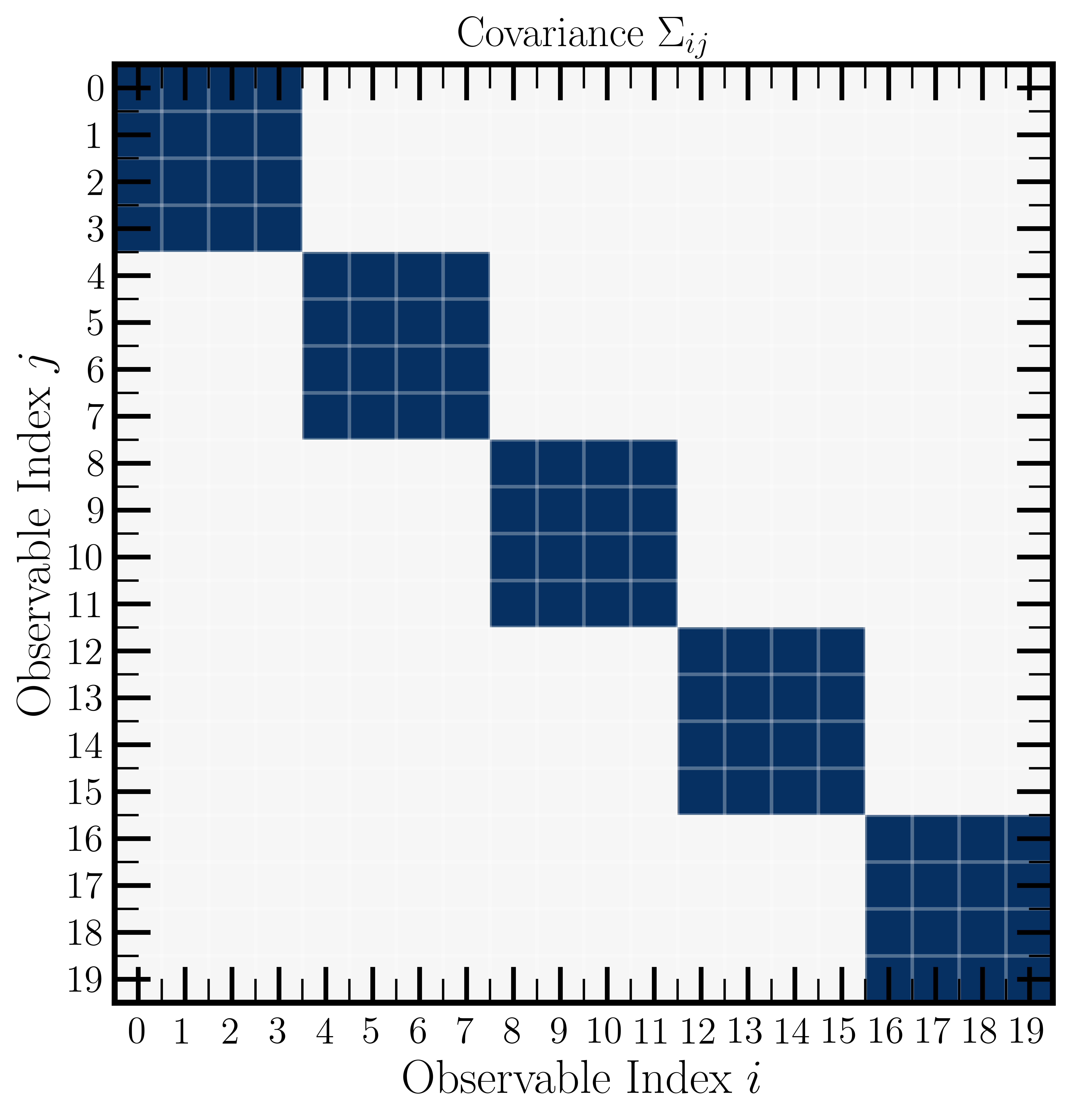}
\includegraphics[width=0.21\textwidth]{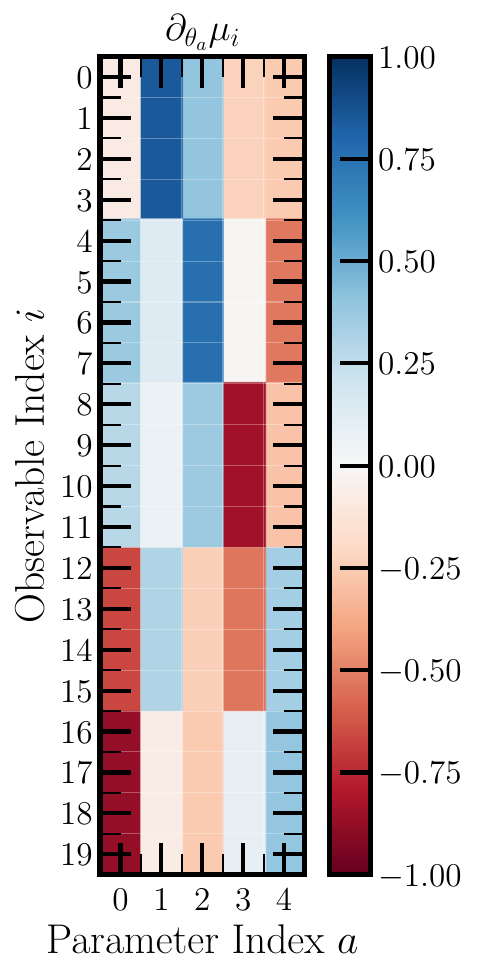}
\caption{(Left) Covariance matrix $\Sigma$ corresponding to the toy example outlined in \cref{eq:multivar_gaussian}, with $K = 20$. The covariance is in block-diagonal form. (Right) The dependence of the means $\mu_i$ of each observable on each of the 5 parameters $\theta_i$. 
}
\label{fig:covariance}
\end{figure}

In this example, a total of $N_\text{obs}= 20$ observables are considered, divided into four identical copies of five distinct observables.
Observables are indexed such that within the sets (0--3)$_0$, (4--7)$_1$, (8--11)$_2$, (12--15)$_3$, and (16--19)$_4$, where the subscript $\alpha$ indexes a set, the observables are identical.
The structure of the corresponding covariance matrix $\Sigma$ is shown in \cref{fig:covariance}~(left).
For simplicity, the means $\mu_i$ are linear combinations of $N_\text{par}=5$ parameters $\theta_a$, so that $\partial \mu_i/\partial {\theta_a}$ is a constant $20 \times 5$ matrix.
We choose (without loss of generality) for $\partial \mu_i/\partial {\theta_a}$ to have a vector norm of 1 across the parameter index, but with otherwise random entries respecting the block-diagonal structure, meaning that within each set $\alpha$ the means $\mu_i$ have identical dependence on parameters $\theta_a$, as shown in \cref{fig:covariance}~(right).

Next, we use the \method score to select the ``best'' set of $K=5$ observables out of these 20.
For various choices of $\beta$, including the heuristic $\beta$ of \cref{eq:heuristic}, we first compute the \method score $\mathcal{S}_\text{HD}$ in \cref{eq:bounded-score} for all ${20 \choose 5}$ combinations of observables, and keep the set with the highest score.
In this toy model, we can compute all single-observable Fisher information matrices since we know the likelihood is Gaussian, and we do not have to make use of the binned techniques of \Sec{single:obs:Fisher}.
In this example, it is clear that the optimal observable set is the one that includes exactly one observable from each of the five sets of observables (it does not matter which observable within each set, as long as all sets are represented).
Anything other than this choice is a suboptimal combination.
Finally, due to the fact that we have constrained all variances to be 1, and $\partial \mu_i/\partial {\theta_a}$ to have a norm of 1,  the trace of the \emph{full} Fisher information matrix is a constant, 
\begin{align}
    \Tr[I_{\rm full}] = 5.0,
\end{align}
by construction. 
We can use this to verify the inequality in \cref{eq:Spr:SHD:ineq}, that the \method score is an approximate lower bound on the trace of the full Fisher information matrix.

\begin{table}[]
    \centering
    \begin{tabular}{c l l l}
        \hline    \hline
        Penalty $\beta$ & Selected observables & ${\mathcal S}_\text{HD}$ score & Quality\\
         \hline  
         -0.1 & \{0$_0$, 1$_0$, 2$_0$, 3$_0$, 16$_4$\} & 6.749 & Maximally subopt. \\
         0.0 & \{0$_0$, 1$_0$, 9$_2$, 10$_2$, 11$_2$\} & 5.000 & Suboptimal \\
         0.1 & \{0$_0$, 4$_1$, 8$_2$, 12$_3$, 16$_4$\} & 4.234 & Optimal \\
         \textbf{0.143} & \textbf{\{0$_0$, 4$_1$, 8$_2$, 12$_3$, 16$_4$\}} & \textbf{3.960} & \textbf{Optimal} \\
         1.0 & \{0$_0$, 4$_1$, 8$_2$, 12$_3$, 16$_4$\} & -2.654 & Optimal\\
      \hline    \hline       
    \end{tabular}
    \caption{The observable selections for the toy Gaussian example for various choices of $\beta$. The bolded row uses the heuristic choice for $\beta$, \cref{eq:heuristic}, and the subscripts index the observable set $\alpha$ from which the observable is drawn. }
    \label{tab:toy_gaussian}
\end{table}

The results of this exercise are summarized in \Tab{toy_gaussian}.
We first note that choosing any positive $\beta$ results in an optimal observable selection;
the selected observables are always independent (\ie, from different ``blocks'').
If $\beta = 0$, this is no longer the case, as some observables share a block while others do not. 
Finally, if $\beta < 0$ (which is ``unphysical''), there is a preference for as many observables as possible to be in the \emph{same} block, which is the \emph{worst} choice one could pick.
This aligns with our intuition that $\beta$ corresponds to a ``correlation penalty''---a positive $\beta$ selects for uncorrelated observables, a negative $\beta$ selects for correlated observables, and $\beta = 0$ lies in between.
While for a given model (a given choice of $\partial \mu_i/\partial {\theta_a}$), the observable selection is completely deterministic, across many random models the $\beta = 0$ score selects for uniformly random sets of observables.

We also note that for all physical $\beta$, the \method score  indeed  is below 5.0---the ``true'' trace of the Fisher information matrix.
Choosing $\beta$ too high, well above the heuristic prescription in \cref{eq:heuristic}, however, can result in a negative score. 
The negative score does not necessarily imply that the observable combination is suboptimal (indeed, in our toy example it is still the optimal set).
Rather, a negative $\mathcal{S}_{\rm HD}$ only implies that the estimated overlap is too high to represent a valid Fisher information trace.
The fact that an overly large $\beta$ chooses the correct observables anyways is due to the simplicity of this example: all observables have the same amount of individual information, and either observables have zero correlation, or they are completely correlated. 
Therefore, increasing $\beta$ above zero will not change the relative ordering of the observable sets. 
This feature is not expected to hold in practical settings.

%%%%%%%%%%%%%%%%%%%%%%%%%%%%%%%%%%%%%%%%%%%%%
\section{Application to Lund string hadronization}
\label{sec:application}

We now demonstrate the use of \method in a realistic setting by applying it to constrain five parameters of the Lund string hadronization model in \pythia~8.3~\cite{Bierlich:2022pfr}. We focus on $e^+e^- \to Z \to \text{jets}$ collisions at $\sqrt{s} = 91.2$ GeV, a clean experimental environment where hadronization effects are well-studied. The five parameters are\footnote{Note that some hadronization parameters are dimensionful (\eg, $\sigma_{p_T}$ set by {\tt StringPT:sigma} or $m_\text{join}$ set by {\tt FragmentationSystems:mJoin} in \pythia). For such parameters, we use logarithmic coordinates, \eg, $\theta_a = \log(\sigma_{p_T})$, to ensure well-behaved gradients across different scales. Logarithmic coordinates are also advised for dimensionless parameters spanning large hierarchies. Parameters that can be negative should be used without transformation.
}
\beq
\label{eq:bmtheta}
\bm \theta = \{\log a_{\rm Lund},\log b_{\rm Lund},\log( \sigma_{p_T}/\text{GeV}),\log \rho, \log \xi\},
\eeq
where:
\begin{itemize}
    \item $a_{\rm Lund}$ and $b_{\rm Lund}$ parametrize the Lund string fragmentation function that determines the lightcone momentum fraction $z$ of produced hadrons:
    \beq
      f(z) \propto \frac{(1-z)^{a_{\rm Lund}}}{z}\exp\left(-\frac{b_{\rm Lund} m^2_{T}}{z}\right).\label{eq:f_of_z}
    \eeq 
    These are set by {\tt StringZ:aLund} and {\tt StringZ:bLund} in \pythia.
    
    \item $\sigma_{p_T}$ controls the transverse momentum distribution of hadrons via Gaussian-distributed kicks at each string break:
    \beq
    P(\Delta p_x,\Delta p_y) = \mathcal{N}\left(0,\frac{\sigma_{p_T}^2}{2}\mathbb{I}_{2\times 2}\right).
    \eeq
    This is set by {\tt StringPT:sigma}, where $m_T^2 = p_T^2 + m^2_{\rm{had}}$  in Eq. \eqref{eq:f_of_z}.
    
    \item $\rho$ controls the suppression of strange quark-antiquark pairs ($s\bar{s}$) relative to light quarks ($u\bar{u}$ or $d\bar{d}$) in string breaks, set by {\tt StringFlav:probStoUD}.
    
    \item $\xi$ controls the suppression of diquark-antidiquark pairs relative to quark-antiquark pairs in string breaks, set by {\tt StringFlav:probQQtoQ}.
\end{itemize}
These parameters capture distinct aspects of the hadronization process: $a_{\rm Lund}$ and $b_{\rm Lund}$ govern longitudinal fragmentation, $\sigma_{p_T}$ controls transverse structure, while $\rho$ and $\xi$ determine flavor composition. This diversity allows us to test whether \method appropriately identifies observables sensitive to different physical effects.

We validate \method through several complementary studies using simulated data. First, we summarize the dataset and observable computations used in the studies (\cref{sec:numerical_details}). Next, we compare \method selections against optimal choices obtained from machine-learning approximations of the full likelihood (\cref{sec:validation}). We then examine the performance of \method across different observable types, including both jet-level and event-level measurements (\cref{sec:observable_types}). Finally, we discuss how detector effects can be incorporated and how the \method score enables meaningful comparisons across different experimental setups (\cref{sec:multiple_experiments,sec:detectors}).

\subsection{Dataset and observables}
\label{sec:numerical_details}

We generate $N=10^6$ events of $e^+e^- \to Z \to q\bar{q}$ at $\sqrt{s} = 91.2$ GeV using \pythia~8.3, where the $Z$ boson decays to all five light quark flavors $(u,d,s,c,b)$. All $N_\text{par}=5$ \pythia parameters, \cref{eq:bmtheta}, are set to their default values, corresponding to the reference point
\beq
\bm{\hat \theta} = \log \{0.68, 0.98, 0.335, 0.217, 0.081\}. 
\eeq
We compute the following $N_\text{obs}=15$ hadronization-sensitive observables, which we define in detail in \cref{sec:observable_definition}:
\beq
\begin{split}
\label{eq:list:O}
\bm{\mathcal{O}} = \big\{&n_{\rm had}, n_{\rm ch}, n_{\rm baryon}, n_{\rm str}, 1-T, B_T, B_W, C, D, 
\\
&n_{\rm jet}, \tau_1, \tau_2, \tau_3, \text{EEC}, \text{NNC}\big\}.
\end{split}
\eeq
These include event-level multiplicity observables ($n_{\rm full}$ through $n_{\rm Strange}$), event shapes ($1-T$ through $D$), jet-based observables (the event-level jet multiplicity $n_{\rm jet}$ and the jet-level $\tau_1$ through $\tau_3$), and sample-level correlation functions (EEC and NNC). 

\paragraph{Binning strategy.}
For each observable, we construct a histogram with bin-width designed to ensure sufficient statistics per bin.
For the angular observables EEC and NNC, we use uniform angular bins of width $\Delta\theta = \pi/50$, matching the resolution of LEP measurements~\cite{AKRAWY1990159}.
For each other observable, we require that each bin contains enough events such that the fractional bin occupancy $\alpha_m = n_m/N$ has a relative statistical uncertainty of at most $1\%$. To achieve this, we use quantile binning with $B = \text{round}(10^{-4}N')$ bins, where $N'$ is the total number of jets in the sample in the case of jet-based observables and is the total number of events $N$ otherwise, which gives approximately $N'/B$ entries per bin and thus $\Delta\alpha_m/\alpha_m \approx \sqrt{B/N'} = 1\%$. Bins with fewer counts are merged with their left neighbor. 

\paragraph{Computing Fisher information matrices.}
As described in \cref{sec:single:obs:Fisher},
the calculation of single observable Fisher information matrices $ I^{(i)}_{ab}({\bm \theta})$ (\cref{eq:Iab:def})
     involves computing gradients $(\partial_{\bm\theta}\alpha_m)|_{\bm{\hat \theta}}$ of the bin occupancies with respect to the hadronization parameters (\cref{eq:Iab:chain:rule}). We compute these gradients by generating alternative parameter points $\bm\theta'_k$ near $\bm{\hat\theta}$, reweighting the original sample to each $\bm\theta'_k$, and then fitting a linear model to the resulting $\alpha_m(\bm\theta'_k)$ values.

To ensure accurate reweighting and stable gradient fits, we sample $N_{\theta'}=150$ parameter points on a hypersphere of radius $5\%$ around the reference point: 
\beq
\exp(\bm\theta'_k) = (1 + 0.05\,\bm\epsilon_k) \cdot \exp(\bm{\hat \theta}), \quad \bm\epsilon_k \in \{\bm{x} \in \mathbb{R}^5 : \|\bm{x}\| = 1\},\quad k=1,\ldots, N_{\theta'},
\label{eq:hypersphere}
\eeq
with the additional constraint 
$\sigma_{p_T}' \leq \hat \sigma_{p_T}$, to ensure good reweighting coverage. The $5\%$ radius balances two competing requirements. 
The parameters must be chosen close enough to ensure the accuracy of the reweighting procedure and the validity of the linear approximation,
$\alpha_m(\bm\theta') \approx \alpha_m(\bm{\hat \theta}) + (\partial_{\bm\theta}\alpha_m) \cdot (\bm\theta' - \bm{\hat \theta})$,
yet sufficiently far apart that numerical precision does not dominate the fitted gradients.
The gradients are obtained by least-squares regression using {\tt scikit-learn}~\cite{scikit-learn}.

%%%%%%%%%%%%%%%%%%%%%%%%%%%%%%%%%%%%%%%%
\subsection{Validation against approximate likelihood}
\label{sec:validation}

To validate the \method score, we compare its selections against an approximate ``gold standard'' based on the full likelihood. Since computing the exact full likelihood for all observable combinations is intractable, we construct an approximation using machine learning: for each subset of $K$ observables, we train a classifier to distinguish events generated at $\bm{\hat \theta}$ from those at a nearby parameter point $\bm\theta'$. The classifier output provides a one-dimensional summary statistic that captures all information from the $K$ observables about the parameters, including their correlations. This approximates the full likelihood-based selection, as per the Neyman-Pearson lemma, allowing us to quantify the loss of  information in the \method score, which treats correlations only approximately.

The validation procedure proceeds as follows for each subset of $K$ observables from our pool of $N_{\text{obs}} = 15$:
\begin{enumerate}
    \item Train a gradient boosting classifier using \textsc{XGBoost}~\cite{Chen_2016} to distinguish baseline events ($\bm{\hat \theta}$) from reweighted events for a single representative $\bm\theta'$, chosen as the parameter point in the sample in \cref{eq:hypersphere} with maximal $|1-\bar w|$ (here, $\bar{w}$ is the mean event weight calculated over the sample of $w_k(\bm{\hat \theta}, \bm \theta')$ in \cref{eq:alpha_m:theta'}). The observables in each event are binned in order to contain the same information as \method, and thus can take only discrete values given by their binned index. Since we have tabular data, Boosted Decision Trees are a natural choice of classifier.
    \item Use the classifier score $s(\mathcal{O}_1,\ldots,\mathcal{O}_K)$ as a one-dimensional summary statistic. Compute its Fisher information matrix $\hat{I}_{\text{full}}$ using the same variable-width binning strategy described in \cref{sec:numerical_details}. 
    \item Invert $\hat{I}_{\text{full}}$ to obtain the Cramér-Rao lower bound on parameter uncertainties via \cref{eq:cov:CR}. Use $\text{Tr}(\hat{I}_{\text{full}}^{-1})$ and $\det[\hat{I}_{\text{full}}^{-1}]$ as figures of merit: smaller values indicate better constraining power. (See \cref{app:sec:determinant:score} for a discussion on the relative merits of the trace and determinant.)
\end{enumerate}

\begin{figure}[h!]
\centering
\includegraphics[width=0.425\linewidth]{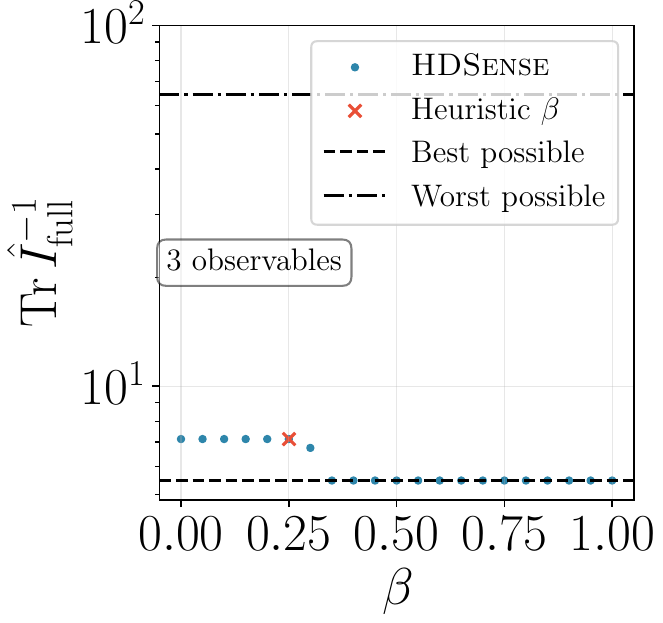}
\includegraphics[width=0.425\linewidth]{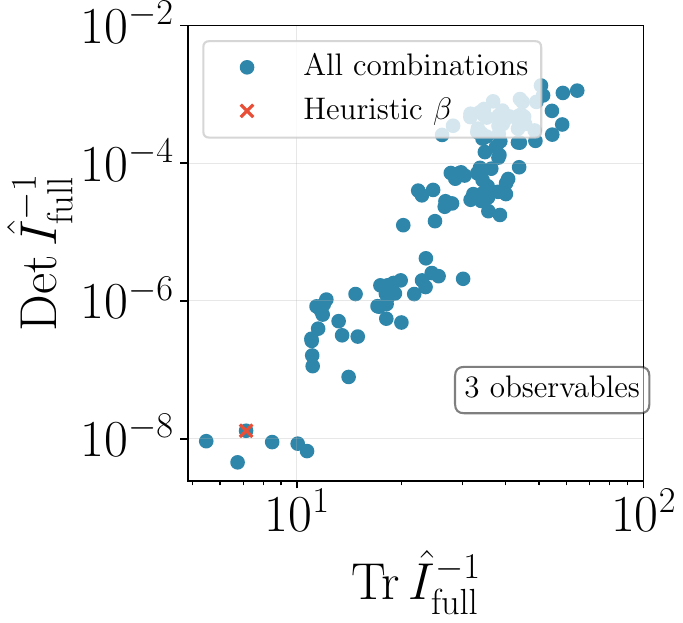}\\
\includegraphics[width=0.425\linewidth]{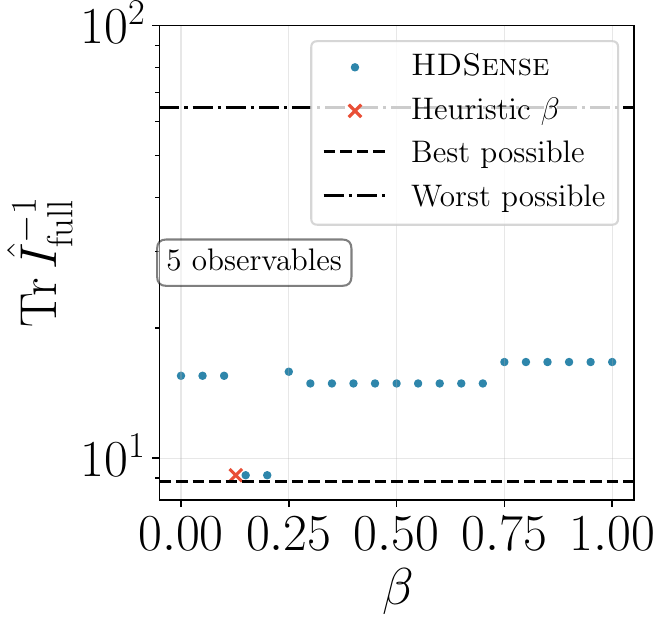}
\includegraphics[width=0.425\linewidth]{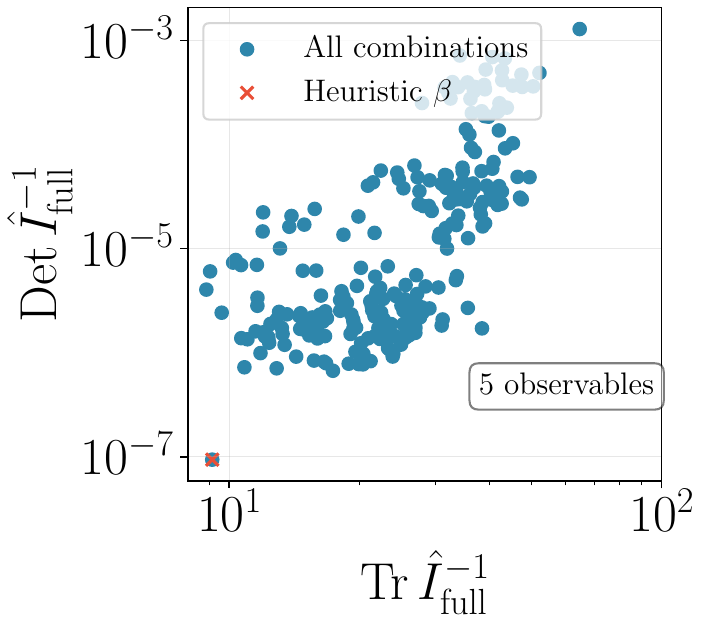}\\
\includegraphics[width=0.425\linewidth]{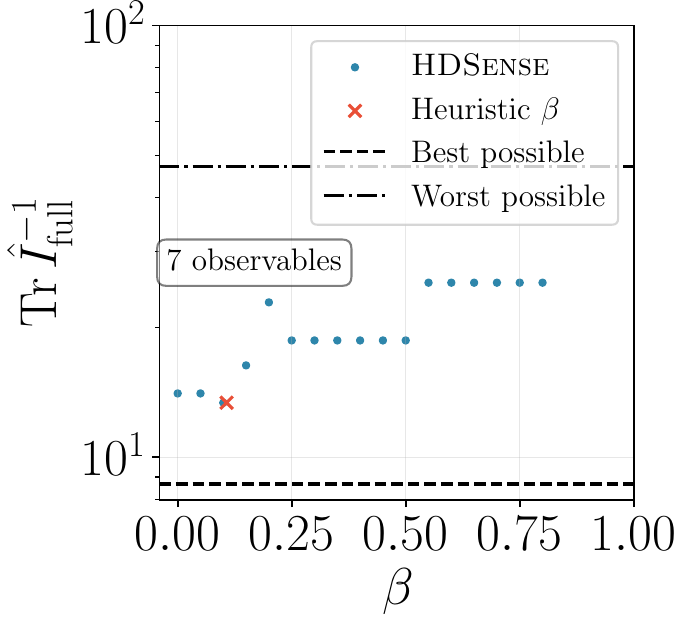}
\includegraphics[width=0.425\linewidth]{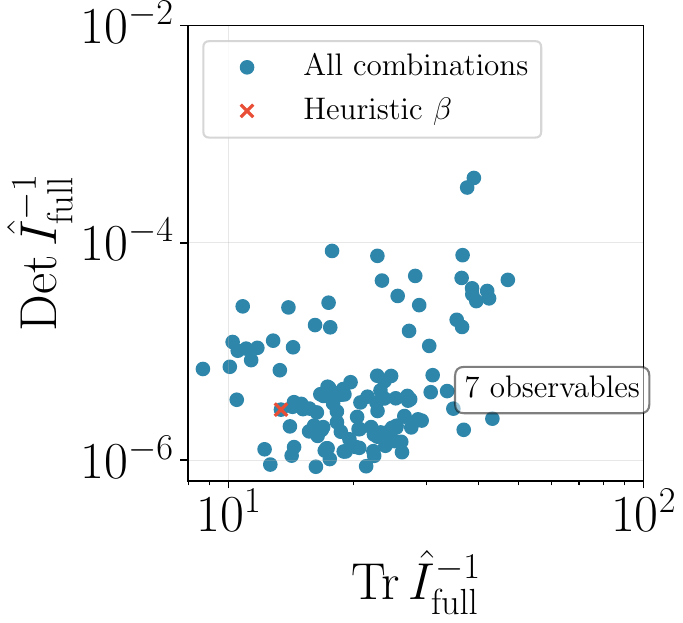}
\caption{
Validation of \method selection prescription against approximate full likelihood for $K = 3$ (top), $K = 5$ (middle), and $K = 7$ (bottom) observables.
\textbf{Left:} Dependence of the full Fisher information trace on $\beta$ for the subset selected by \method at each $\beta$ value. The smallest and largest possible traces, obtained by exploring all possible $K$-observable subsets, are shown for reference.
\textbf{Right:} Trace versus determinant of $\hat{I}_{\text{full}}^{-1}$ for all possible $K$-observable subsets. Better subsets lie closer to the origin.
Red cross shows the heuristic choice $\beta = 0.5/\max\mathcal{P}_{\text{overlap}}$.
} 
\label{fig:scatter_and_scan}
\end{figure}

\Cref{fig:scatter_and_scan} shows the validation results for $K = 3, 5, 7$ observables. The left column examines the sensitivity of \method to the hyperparameter $\beta$, plotting $\text{Tr}(\hat{I}_{\text{full}}^{-1})$ for the subset selected by \method as a function of $\beta$. The red cross marks our heuristic choice $\beta = 0.5/\max_{\mathcal{X}}\mathcal{P}_{\text{overlap}}(\mathcal{X})$ (\cref{eq:heuristic}). The right column shows $\text{Tr}(\hat{I}_{\text{full}}^{-1})$ versus $\det[\hat{I}_{\text{full}}^{-1}]$ for all possible $K$-observable subsets. Optimal selections lie near the origin (small trace and determinant). The red cross indicates the \method selection.

\begin{figure}[tb]
\centering
\includegraphics[width=0.65\linewidth]{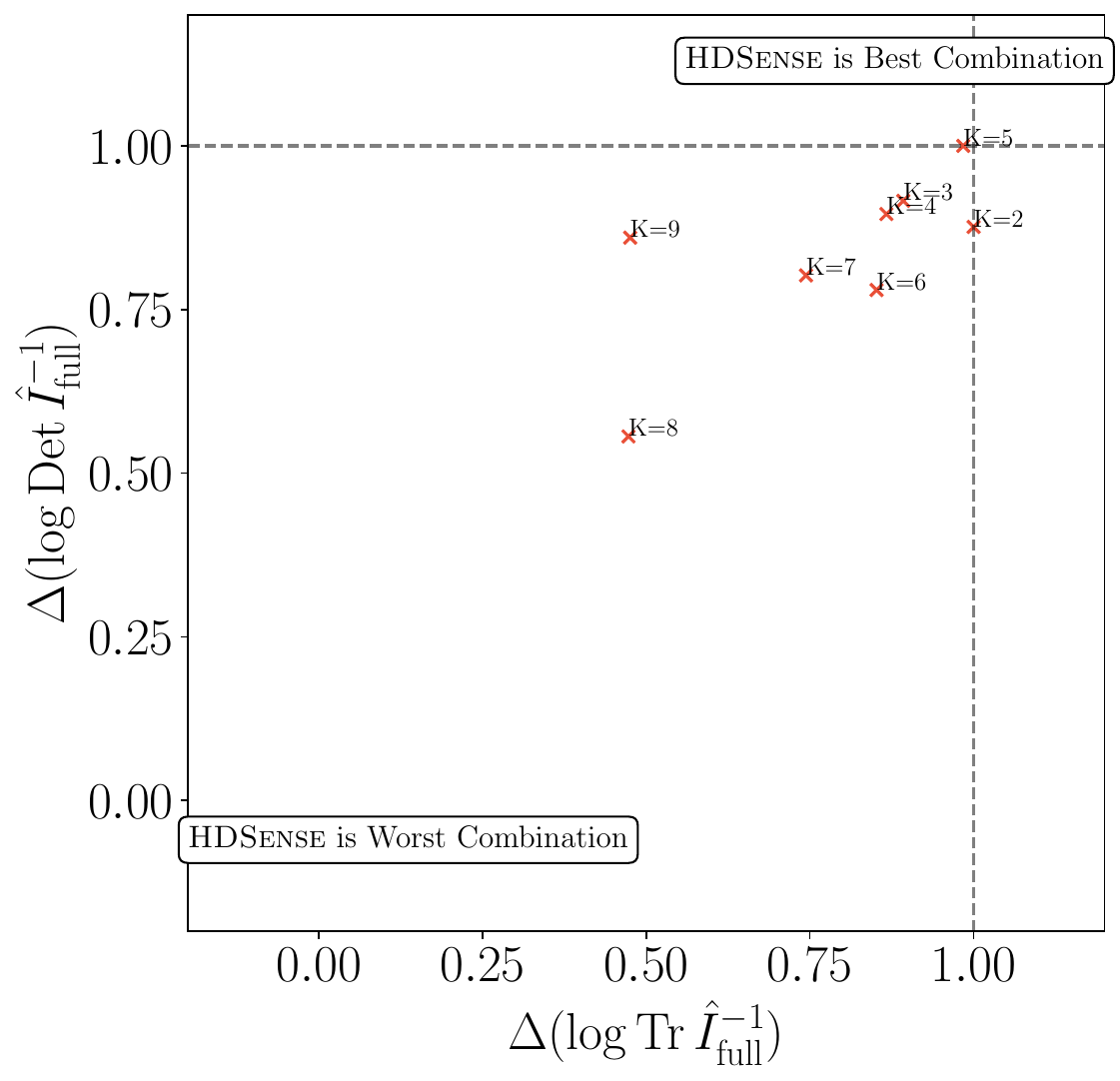}
\caption{Comparison of the selected subsets for all $K$. For $K\leq 7$, the \method score hovers close to the optimal region, while it remains far from the `worst' possible score for all $K$. The quality of the approximation degrades for large $K$ where correlations become more important. 
Here, the $\Delta$ metric is defined as the ratio between the difference in log space of the worst combination and the combination selected by \method and the difference in log space between the best and worst combinations. All information matrices are computed using the approximate full likelihoods.
} 
\label{fig:temp_comparison}
\end{figure}

Several features are evident. First, the heuristic choice of $\beta$ consistently lies in a reasonable regime: while not always optimal, it avoids extreme values that would either ignore correlations entirely ($\beta \to 0$) or over-penalize them ($\beta \to \infty$). Second, the \method selections (right column, red crosses) hover near the optimal region, particularly for small $K$ where correlations between observables are less important (see the discussion of the penalty term in \cref{sec:hd_sense}). For $K = 3,5$, \method performs nearly optimally. As $K$ increases, the selection task becomes more challenging: with many observables, one must balance the trace (dominated by the largest eigenvalue of the covariance matrix, related to the worst-constrained parameter direction) against the determinant (the covariance volume in parameter space). \method provides a reasonable compromise in this trade-off, consistently selecting competitive subsets even when correlations become more significant at larger $K$. We compare the performance across $K$ in \cref{fig:temp_comparison}, where we observe how for $K\leq 7$, the selected observable subset is close to the optimal subset as measured by the $\Tr[\hat{I}_{\rm full}]$ and $\det[\hat{I}^{-1}_{\rm full}]$ around the best selection, while there is some slight degradation for $K=8,9$.

This validation demonstrates that \method, despite its computational simplicity and ignorance of inter-observable correlations, makes selections that approximate those from the full likelihood.  The method is particularly effective when $K$ is small relative to the number of parameters, where overlap penalties naturally guide the selection toward complementary observables. Note that the selection of the ``optimal'' set of observables also carries an inherent ambiguity if the full likelihood is used, since the selections based on trace or on determinant ($A$-optimality-like or $D$-optimality-like, as discussed in \cref{app:sec:determinant:score}) of $\hat{I}_{\text{full}}^{-1}$ often differ, while \method strikes a balance between the two criteria. 

%%%%%%%%%%%%%%%%%%%%%%%%%%%%%%%%%%%%%%%%%%%
\subsection{Observable selection and ranking}
\label{sec:observable_types}

\begin{figure}[t!]
\centering
\includegraphics[width=0.9\textwidth]{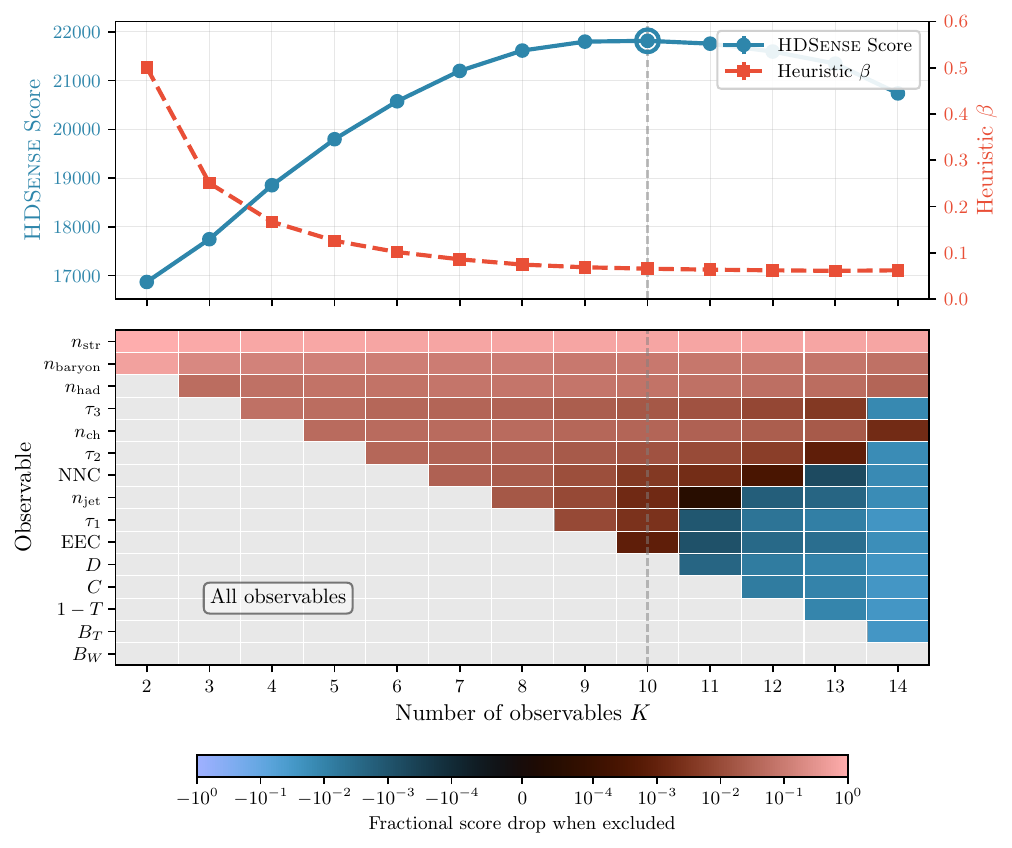}
\caption{\method ranking for all 15 observables using the exhaustive search algorithm, as assessed across bootstrapped samples (see text).
\textbf{Top:} \method score (blue) and heuristic $\beta$ (red) versus number of selected observables $K$. The maximum \method score is highlighted with an additional blue circle and a dashed vertical line is drawn at the corresponding $K$-value.
\textbf{Bottom:} Heatmap showing the fractional score reduction $\Delta\mathcal{S}_{\text{HD}}/\mathcal{S}_{\text{HD}}$ when each observable is excluded from the selected subset. 
% Darker colors indicate higher importance. 
Gray cells denote observables not yet selected at that $K$. Observables are ordered by their entry into the greedy selection algorithm when reducing from $K$ to $K-1$. 
A dashed vertical line is drawn at the $K$-value corresponding to the maximum \method score.
}
\label{fig:all_observables_no_detector}
\end{figure}

\begin{figure}[t!]
\centering
\includegraphics[width=0.9\textwidth]{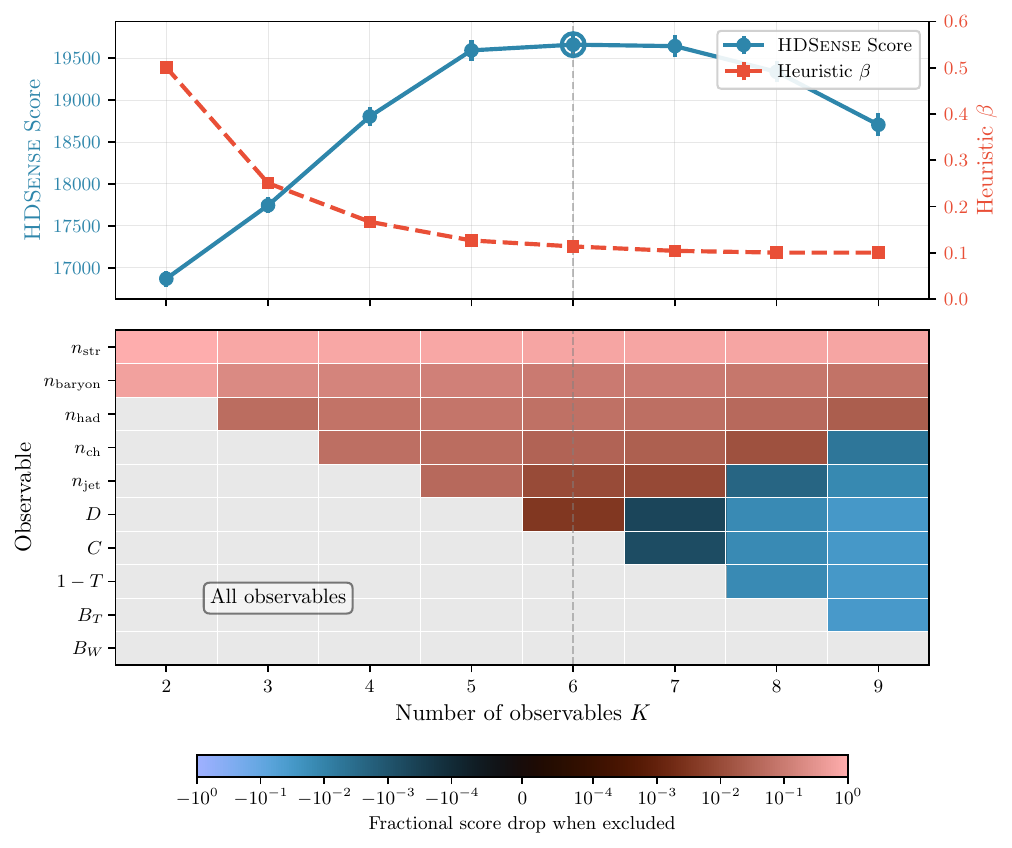}
\caption{\method ranking using only the 9 event-level observables. Layout and interpretation as in \cref{fig:all_observables_no_detector}.
}
\label{fig:event_level_no_detector}
\end{figure}

Having validated \method against the approximate full likelihood, we now examine its selections in detail. \Cref{fig:all_observables_no_detector} shows the \method ranking for all 15 observables (event-level, jet-level, and ensemble-level), while \cref{fig:event_level_no_detector} shows results using only the subset of 9 event-level observables. Both figures display two panels: the top panel shows the \method score and heuristic $\beta$ as functions of the number of selected observables $K$ using the exhaustive search algorithm, while the bottom panel shows a heatmap indicating the importance of each observable within each selected subset.

To assess the stability of \method selections, we bootstrap the dataset 100 times and recompute rankings for each bootstrap sample. We treat event-level, jet-level, and ensemble-level observables separately due to their different statistical structures. For event-level and jet-level observables, each bootstrap sample is constructed by: (1) drawing the total number of events (or jets) from a Poisson distribution with mean equal to the original sample size, then (2) resampling that many events (or jets) with replacement from the original dataset, and (3) recomputing histograms with adaptive binning as described in \cref{sec:numerical_details}. Event weights are preserved in the resampling, so reweighted histograms for different $\bm\theta$ values are obtained directly from the weighted bootstrap samples. For ensemble-level observables (EEC and NNC), which are already binned distributions, we resample histogram bins directly: for each parameter point $\bm\theta$, we draw new bin values from a multivariate Gaussian with mean and covariance determined by the original weighted histogram.

The top panel shows the \method score and heuristic $\beta$ averaged over all bootstrap samples, with standard deviations indicated by vertical error bars. The bottom panel requires care to interpret, because different bootstrap samples may select different observables at fixed $K$. For each $K$, we identify those observables most frequently selected across all bootstrap samples and display the top $K$. 

The heatmap visualizes the fractional score drop when each observable is removed from the selected $K$-observable subset. Specifically, the fractional drop for observable $i$ at subset size $K$ is:
\beq
\frac{\Delta\mathcal{S}_{\text{HD}}^{(i)}}{\mathcal{S}_{\text{HD}}} = 1 - \frac{\langle\mathcal{S}_{\text{HD}}(\mathcal{X}_K \setminus \{i\})\rangle_{i \in \mathcal{X}_K}}{\langle\mathcal{S}_{\text{HD}}(\mathcal{X}_K)\rangle_{\text{all}}},
\eeq
where the numerator averages over only those bootstrap runs where observable $i$ was actually selected in $\mathcal{X}_K$, while the denominator averages over all runs. Gray cells correspond to observables not yet selected at that $K$. Observables are ordered vertically by their entry into the greedy selection: those appearing first (top rows) are selected earliest and typically maintain high importance throughout. We emphasize that this greedy selection is solely used to determine the observable ordering, and that the actual selection is exhaustive.

Several patterns emerge from these results. First, \method prioritizes infrared and collinear (IRC)-unsafe observables such as multiplicities ($n_{\text{had}}$, $n_{\text{ch}}$, $n_{\text{baryon}}$, $n_{\text{str}}$), which enter the selection early and maintain strong importance. An observable is IRC-safe if it remains unchanged under soft emissions or collinear splittings~\cite{Banfi:2004nk}; multiplicities violate both conditions and are thus IRC-unsafe. These multiplicities are directly sensitive to the flavor parameters $\rho$ and $\xi$ while also being indirectly influenced by the other hadronization parameters. In contrast, IRC-safe event shapes, such as thrust, are selected only for higher $K$ values, since their information partially overlaps with that contained in the multiplicities.

Second, the heuristic $\beta$ decreases with increasing $K$, reflecting that correlations become more significant as more observables are added. This adaptive $\beta$ has the beneficial property that the selected subset grows monotonically: the $K$-observable subset is always contained within the $(K+1)$-observable subset. This monotonicity is not guaranteed for fixed $\beta$, as demonstrated in \cref{app:fixed_beta}.

Third, the \method score exhibits non-monotonic behavior, reaching a maximum around $K = 10$ in \cref{fig:all_observables_no_detector}. This does not indicate that $K = 10$ is optimal; rather, it reflects a limitation of the heuristic $\beta$ prescription, which may over-penalize correlations for large $K$. As shown in \cref{app:fixed_beta}, the score with fixed $\beta$ continues to increase with $K$, confirming that the non-monotonicity arises from our correlation approximation rather than from genuine information saturation. The choice of $K$ should be guided by experimental constraints (resources, time, personpower, etc), and maximizing the \method score should not be a criterion in selecting $K$.

%%%%%%%%%%%%%%%%%%%%%%%%%%%%%%%%%%%%%%%%%%%%%%%%%%%%%
\subsection{Combining measurements across experiments}
\label{sec:multiple_experiments}

An advantage of \method score is that it accommodates measurements from multiple experiments with different event statistics and observable coverage. This is particularly relevant for hadronization studies, where different collider experiments may measure complementary sets of observables with varying precision.

When combining uncorrelated measurements from $M_{\text{exp}}$ ideal experiments (we defer detector effects to \cref{sec:detectors}), the single-observable Fisher information matrices generalize straightforwardly from \cref{eq:Iab:chain:rule},
\beq
 I_{ab}^{(i)}(\bm \theta) = \sum_{\gamma=1}^{M_{\text{exp}}} I_{ab}^{(i,\gamma)}(\bm \theta),
\label{eq:multi_exp_fisher}
\eeq
where $I_{ab}^{(i,\gamma)}$ is the contribution from experiment $\gamma$. If all experiments measure the same set of bins with perfect efficiency, the contributions scale simply with the number of events $N_\gamma$ collected by each experiment,
\beq
 I_{ab}^{(i,\gamma)}(\bm \theta) = \frac{N_\gamma}{N} I_{ab}^{(i)}(\bm \theta),
\label{eq:n_gamma_info}
\eeq
where $N = \sum_\gamma N_\gamma$ is the total event count. This reflects the basic principle that Fisher information is additive for independent measurements: larger datasets provide proportionally more constraining power.

To illustrate the interplay between statistics and observable coverage, we consider two experiments:
\begin{itemize}
    \item \textbf{Experiment I}: Collected $N_{\text{I}} = 10^6$ events and measured all event-level observables except $n_{\text{baryon}}$ and $n_{\text{str}}$ (no particle identification capability).
    \item \textbf{Experiment II}: Collected $N_{\text{II}} = 10^5$ events ($10\times$ fewer than Experiment I) but with full particle identification, enabling measurement of all event-level observables including $n_{\text{baryon}}$ and $n_{\text{str}}$.
\end{itemize}
In computing the \method score via \cref{eq:bounded-score}, most observables use the combined sample size $N = 1.1 \times 10^6$ events, while $n_{\text{baryon}}$ and $n_{\text{str}}$ use only $N = 10^5$ events from Experiment II.

\begin{figure}[t!]
\centering
\includegraphics[width=0.9\textwidth]{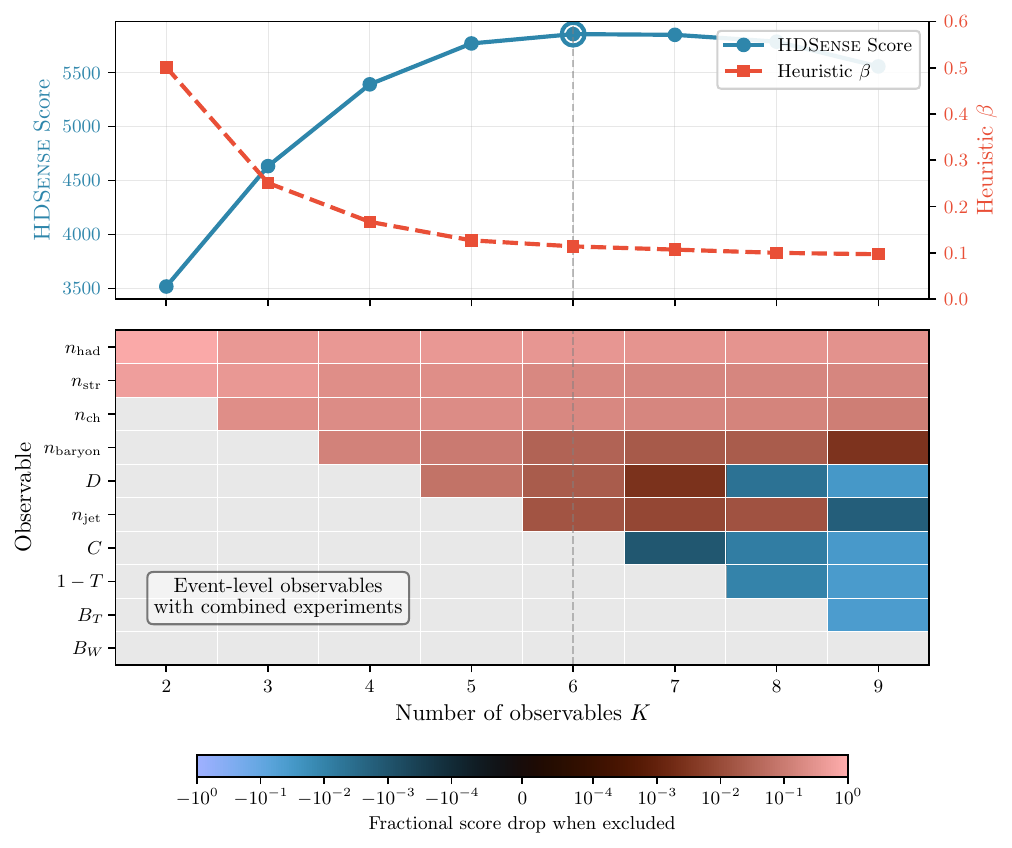}
\caption{\method ranking for event-level observables when combining two experiments with different statistics and observable coverage. Experiment I: $10^6$ events, no particle ID (cannot measure $n_{\text{baryon}}$, $n_{\text{str}}$). Experiment II: $10^5$ events, full particle ID. 
\textbf{Top:} \method score (blue) and heuristic $\beta$ (red) versus $K$.
\textbf{Bottom:} Observable importance heatmap, with layout as in \cref{fig:all_observables_no_detector}. Despite $10\times$ lower statistics, $n_{\text{baryon}}$ and $n_{\text{str}}$ maintain high ranking due to their unique sensitivity to flavor parameters.
}
\label{fig:event_level_different_experiments}
\end{figure}

\Cref{fig:event_level_different_experiments} shows the resulting \method ranking. Despite having $10\times$ lower statistics, the baryon and strangeness multiplicities from Experiment II remain highly ranked. This occurs because they are the only observables directly sensitive to the flavor parameters $\rho$ and $\xi$, making them irreplaceable despite their statistical disadvantage. Meanwhile, the other multiplicities ($n_{\text{had}}$, $n_{\text{ch}}$) gain increased prominence due to the boosted statistics from combining both experiments.

This example demonstrates that \method appropriately balances statistical precision against unique information content: observables with lower statistics can still be selected if they probe parameter directions that other observables cannot access. This is particularly valuable when designing combined analyses across experiments with different capabilities.

%%%%%%%%%%%%%%%%%%%%%%%%%%%%%%%%%%%%%%%%
\subsection{Accounting for detector effects}
\label{sec:detectors}

Real experiments face acceptance limitations, reconstruction inefficiencies, and finite resolution that affect which events and particles are measured. \method naturally incorporates these detector effects through the Fisher information framework.

Detector effects modify the bin occupancy fractions $\alpha_m$ for each observable without changing the mathematical structure of the Fisher information. If experiment $\gamma$ has efficiency $\epsilon_\gamma(p_T, \eta, \text{PID}, \ldots)$ depending on particle kinematics and identification capabilities, this efficiency directly affects which events populate each bin of the observable distributions. The single-observable Fisher information from multiple experiments with different efficiencies is
\beq
\label{eq:Iab:theta:detector:eff}
 I_{ab}^{(i)}(\bm \theta) = \sum_{\gamma=1}^{M_{\text{exp}}} I_{ab}^{(i,\gamma)}(\bm \theta),
\eeq
where each experiment's contribution takes the same form as \cref{eq:Iab:chain:rule}:
\begin{equation}
    I_{ab}^{(i,\gamma)}(\bm \theta) = \sum_{m,n=1}^{B_\gamma-1}\frac{\partial\alpha_{m,\gamma}}{\partial\theta_a} \frac{\partial\alpha_{n,\gamma}}{\partial\theta_b} \tilde I_{mn}^{(\gamma)}(\bm \alpha),
    \label{eq:Iab:chain:rule:multi}
 \end{equation}
with
 \begin{equation}
   \tilde I_{mn}^{(\gamma)}(\bm{\alpha}) = N_\gamma\biggr(\delta_{mn}\frac{1}{\alpha_{m,\gamma}} + \frac{1}{1-\sum_{p}\alpha_{p,\gamma}}\biggr).
    \label{eq:tildeImn:multi}
\end{equation}
Here $N_\gamma$ is the number of events collected by experiment $\gamma$, and $\alpha_{m,\gamma} = \mathbb{E}[n_{m,\gamma}/N_\gamma]$ is the expected fraction of events in bin $m$ for that experiment, accounting for its specific efficiency function.\footnote{Note that, as in \cref{sec:single:obs:Fisher}, we drop the $\cO_i$ label on $\tilde I_{mn}^{(\gamma)}(\bm \alpha)$, $\alpha_{n,\gamma}$, and $n_{m,\gamma}$, to simplify the notation.} Note that different experiments may use different binning schemes (varying $B_\gamma$) to match their resolution capabilities, which is automatically incorporated in the  single-observable Fisher information matrix combining multiple experiments, \cref{eq:Iab:theta:detector:eff}.

The formulation of $I_{ab}^{(i)}(\bm \theta)$ in Eqs.~\eqref{eq:Iab:theta:detector:eff}-\eqref{eq:tildeImn:multi}  follows from the statistical independence of experiments---the joint likelihood factorizes into a product of multinomial distributions, one per experiment:
\begin{equation}
    p(\bm{n}|\bm{\alpha}) = 
    \prod_{\gamma=1}^{M_{\text{exp}}} 
    \biggr(1-\sum_{p=1}^{B_\gamma-1}\alpha_{p,\gamma}\biggr)^{n_{B_\gamma,\gamma}}\prod_{m=1}^{B_\gamma-1}\alpha_{m,\gamma}^{n_{m,\gamma}} .
\end{equation}
The Fisher information is additive because each experiment provides independent information about the parameters.

\begin{figure}[t!]
\centering
\includegraphics[width=0.9\textwidth]{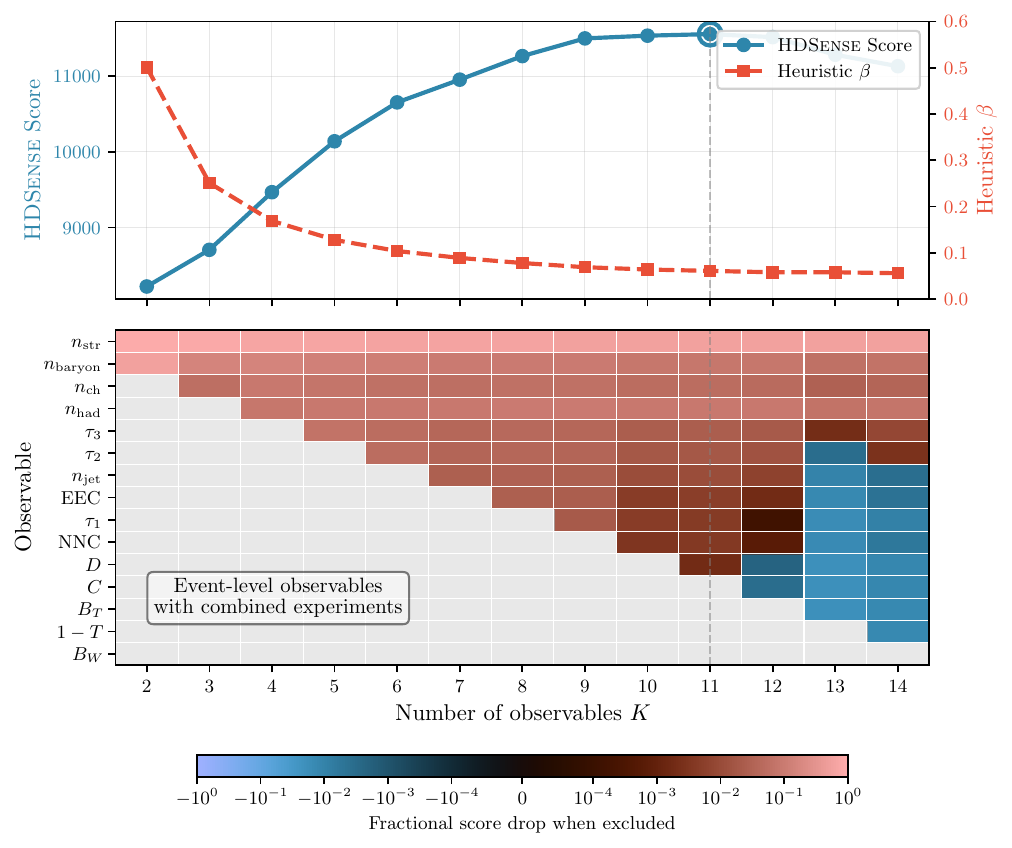}
\caption{\method ranking with estimated detector efficiencies for charged particles as functions of $|p|$ and particle id. Layout as in \cref{fig:all_observables_no_detector}. While overall scores decrease compared to perfect efficiency, the relative ranking of observables remains largely stable.
}
\label{fig:detector_effects}
\end{figure}

Practically, the efficiencies are incorporated by computing $\alpha_{m,\gamma}$ and $\partial\alpha_{m,\gamma}/\partial\theta_a$ using weighted Monte Carlo simulations, as explained in \cref{sec:single:obs:Fisher}, but now with the Monte Carlo simulations that include the detector acceptance and reconstruction effects for experiment $\gamma$. 
To illustrate the impact of detector effects, we consider one such possible efficiency function, inspired by that given in Ref.~\cite{Schyns:1996xrr}. We veto any charged or neutral hadron with $|p|<0.7\,\GeV$ and, for the stable, charged hadrons (namely pions, kaons and protons) that survive and that have $|p|<45\,\GeV$, we consider an additional efficiency function that encodes for the probability that the hadron is correctly tagged as a pion, kaon or proton, or that it is missed. Any hadrons with $|p| > 45\,\GeV$ have perfect efficiency. We consider a benchmark where
\beq
\begin{split}
    p(k|\pi^{\pm})&=\{0.8,0.05,0.05,0.1\}\nonumber\\
    p(k|K^{\pm})&=\{0.2,0.6,0.15,0.05\}\nonumber\\
    p(k|p)&=\{0.2,0.2,0.55,0.05\}\nonumber\\
    \text{where }k &= \{\pi^{\pm},K^{\pm},p,\text{missed}\}\nonumber
\end{split}
\eeq
The effect of applying this efficiency function to all observables, modifying their bin occupancies $\alpha_m$ accordingly, is shown in \cref{fig:detector_effects}.

The results demonstrate that detector limitations primarily affect the absolute Fisher information, reducing the \method score overall when comparing to the relevant case with no detector effects (\cref{fig:event_level_different_experiments} with \cref{fig:event_level_no_detector} and  \cref{fig:detector_effects} with \cref{fig:all_observables_no_detector}, respectively),
but have more modest effects on the relative ranking of observables, which is evidenced also by the heuristic $\beta$ values remaining relatively unchanged. 
Observables that rely heavily on low-$p_T$ or forward particles are somewhat downweighted, but the overall selection pattern remains similar to the ideal case. 
This robustness occurs because \method compares observables under the same experimental conditions: since all observables are computed using the same collection of particles which have undergone detector effects, they all face the same efficiency limitations and so their relative information content is largely preserved.

%%%%%%%%%%%%%%%%%%%%%%%%%%%%%%%%%%%%%%%%%%%
\section{Conclusions}
\label{sec:conclusions}

We have introduced \method, a computationally efficient method for ranking which observables are most valuable for constraining  model parameters with limited information, based on maximizing the trace of the Fisher information, and applied it to the setting of hadronization.
Given a set of candidate observables, \method identifies subsets that maximize information content while avoiding redundancy from correlated measurements. The method requires only single-observable Fisher information matrices, and a heuristic choice for the correlation penalization $\beta$, making it practical for large-scale applications.

We validated \method both with a toy model and by ranking a set of benchmark observables for constraining five Lund string model parameters in $e^+e^- \to Z \to \text{jets}$ collisions. Comparisons to full likelihood approximations show that \method successfully identifies near-optimal observable subsets despite ignoring detailed correlations. For the latter dataset, the rankings favor particle multiplicities over event shapes, reflecting their known sensitivity to flavor parameters. The method also naturally handles multiple experiments with different detector capabilities, balancing statistical power against complementary information.

This work addresses a pressing need in collider physics. Hadronization is the dominant source of systematic uncertainty in many measurements, yet it lacks first-principles predictions and relies on phenomenological models that contain many parameters that must be fit to data, making the choice of which observables to measure critical. \method provides concrete guidance: it informs an experimentalist which measurements will reduce hadronization uncertainties, and informs a phenomenologist on which subsets of observables to base Monte Carlo tuning. This is particularly valuable as LHC experiments plan future analyses and detector upgrades, where resource constraints force difficult prioritization decisions.

Beyond traditional tuning, \method can guide data-driven approaches. Machine learning methods for learning fragmentation functions and neural network hadronization models require training data, and \method identifies which observables provide maximum learning leverage. The framework extends naturally to other domains: parton distribution functions, effective field theory, and any parameter inference problem with limited measurement resources. This also extends beyond particle physics, as in principle these same techniques may also be applied to problems in cosmology, astrophysics, or generic experimental design.
Extending \method to these new problems requires care, however, in the choice of the overlap penalization heuristic $\beta$, which depends implicitly on the quality of the approximation of the complicated correlation structure via \method. 
This may limit the effectiveness of \method when dealing with a very large number of low-level observables (\eg, the raw four-momenta of all particles), where correlations are strong and nonlinear and each observable individually carries only a small amount of information.

Finally, we note that although we derived \method in \cref{sec:derivation} under a Gaussian approximation were correlations were assumed to be parameter-independent, a successful extraction of subsets of observables extends beyond the fully Gaussian case, as shown in \cref{sec:application}. We expect that moving away from either the Gaussian assumption or the assumption of parameter-independent correlations will not induce any pathologies in \method, merely potentially reduce its optimality.

\paragraph{Acknowledgments.} 
We thank Stefan H\"oche, Max Knobbe, Konstantin Matchev, and Prasanth Shyamsundar for useful comments.
BA, PI, RG, and JZ acknowledge support in part by 
DE-SC0026301, and by NSF grants OAC-2103889, OAC-2411215, and OAC-2417682. BA, RG, and JZ acknowledge support in part by DOE grant DE-SC101977.
This manuscript has been co-authored by Fermi Forward Discovery Group, LLC under Contract No. 89243024CSC000002 with the U.S. Department of Energy, Office of Science, Office of High Energy Physics. 
The work of TM is supported in part by the Shelby Endowment for Distinguished Faculty at the University of Alabama and by Fermilab via Subcontract 725339.
This work was performed in part at the Aspen Center for Physics, with support for BA\ by a grant from the Simons Foundation (1161654, Troyer).  The work of MS is partially supported by CONICET.
\section*{Code and Data}

Code implementing \method is available at \url{https://gitlab.com/pythia8-contrib/packages/hdsense}. \method makes use of the \pythia plug-in platform to store all relevant information from simulated events. As detailed in the README of the repository, all the code that generates the necessary events, re-bins the different variables, trains the necessary classifiers and produces all figures can be found in the `share/HDSense/examples/' folder. 

\bibliography{data_paper.bib}

@article{kiefer1959optimum,
  author       = {Kiefer, Jack},
  title        = {{Optimum Experimental Designs}},
  journal      = {Journal of the Royal Statistical Society: Series B (Methodological)},
  volume       = {21},
  number       = {2},
  pages        = {272--304},
  month        = jul,
  year         = {1959},
  doi          = {10.1111/j.2517-6161.1959.tb00338.x},
  url          = {https://doi.org/10.1111/j.2517-6161.1959.tb00338.x}
}

@article{Assi:2025ibi,
    author = {Assi, Beno{\^\i}t and H{\"o}che, Stefan and Lee, Kyle and Thaler, Jesse},
    title = "{QCD Theory Meets Information Theory}",
    eprint = "2501.17219",
    archivePrefix = "arXiv",
    primaryClass = "hep-ph",
    reportNumber = "FERMILAB-PUB-25-0029-T, MIT-CTP 5827, MCNET-25-01",
    doi = "10.1103/gf42-qzd9",
    journal = "Phys. Rev. Lett.",
    volume = "135",
    number = "13",
    pages = "131901",
    year = "2025"
}

@article{Fisher:1925,
    author = "Fisher, R. A.",
    title = "{Theory of Statistical Estimation}",
    journal = "Proc. Cambridge Phil. Soc.",
    volume = "22",
    pages = "700--725",
    year = "1925"
}

@article{Buckley:2009bj,
    author = "Buckley, Andy and Hoeth, Hendrik and Lacker, Heiko and Schulz, Holger and von Seggern, Jan Eike",
    title = "{Systematic event generator tuning for the LHC}",
    eprint = "0907.2973",
    archivePrefix = "arXiv",
    primaryClass = "hep-ph",
    reportNumber = "IPPP-09-52, DCPT-104-22, LU-TP-09-18, HU-EP-09-33, MCNET-09-14",
    doi = "10.1140/epjc/s10052-009-1196-7",
    journal = "Eur. Phys. J. C",
    volume = "65",
    pages = "331--357",
    year = "2010"
}

@article{NNPDF:2021njg,
    author = "Ball, Richard D. and others",
    collaboration = "NNPDF",
    title = "{The path to proton structure at 1{\%} accuracy}",
    eprint = "2109.02653",
    archivePrefix = "arXiv",
    primaryClass = "hep-ph",
    reportNumber = "Edinburgh 2021/12, Nikhef-2021-013, TIF-UNIMI-2021-11",
    doi = "10.1140/epjc/s10052-022-10328-7",
    journal = "Eur. Phys. J. C",
    volume = "82",
    number = "5",
    pages = "428",
    year = "2022"
}

@article{Ellis:2020unq,
    author = "Ellis, John and Madigan, Maeve and Mimasu, Ken and Sanz, Veronica and You, Tevong",
    title = "{Top, Higgs, Diboson and Electroweak Fit to the Standard Model Effective Field Theory}",
    eprint = "2012.02779",
    archivePrefix = "arXiv",
    primaryClass = "hep-ph",
    reportNumber = "KCL-PH-TH/2020-73, CERN-TH-2020-202",
    doi = "10.1007/JHEP04(2021)279",
    journal = "JHEP",
    volume = "04",
    pages = "279",
    year = "2021"
}

@article{Ethier:2021bye,
    author = "Ethier, Jacob J. and Magni, Giacomo and Maltoni, Fabio and Mantani, Luca and Nocera, Emanuele R. and Rojo, Juan and Slade, Emma and Vryonidou, Eleni and Zhang, Cen",
    collaboration = "SMEFiT",
    title = "{Combined SMEFT interpretation of Higgs, diboson, and top quark data from the LHC}",
    eprint = "2105.00006",
    archivePrefix = "arXiv",
    primaryClass = "hep-ph",
    reportNumber = "OUTP-20-05P, Nikhef-2020-020, CP3-21-12, MCNET-21-07,
  MAN/HEP/2021/004",
    doi = "10.1007/JHEP11(2021)089",
    journal = "JHEP",
    volume = "11",
    pages = "089",
    year = "2021"
}

@article{Planck:2018vyg,
    author = "Aghanim, N. and others",
    collaboration = "Planck",
    title = "{Planck 2018 results. VI. Cosmological parameters}",
    eprint = "1807.06209",
    archivePrefix = "arXiv",
    primaryClass = "astro-ph.CO",
    doi = "10.1051/0004-6361/201833910",
    journal = "Astron. Astrophys.",
    volume = "641",
    pages = "A6",
    year = "2020",
    note = "[Erratum: Astron.Astrophys. 652, C4 (2021)]"
}

@book{Cramer:1946,
    author = "Cram\'er, Harald",
    title = "{Mathematical Methods of Statistics}",
    publisher = "Princeton University Press",
    year = "1946"
}

@article{Banfi:2004nk,
    author = "Banfi, Andrea and Salam, Gavin P. and Zanderighi, Giulia",
    title = "{Resummed event shapes at hadron - hadron colliders}",
    eprint = "hep-ph/0407287",
    archivePrefix = "arXiv",
    reportNumber = "FERMILAB-PUB-04-117-T, LPTHE-04-18, NIKHEF-2004-006",
    doi = "10.1088/1126-6708/2004/08/062",
    journal = "JHEP",
    volume = "08",
    pages = "062",
    year = "2004"
}

@article{Rao:1945,
    author = "Rao, C. Radhakrishna",
    title = "{Information and the accuracy attainable in the estimation of statistical parameters}",
    journal = "Bull. Calcutta Math. Soc.",
    volume = "37",
    pages = "81--91",
    year = "1945"
}

@article{Hou:2019efy,
    author = "Hou, Tie-Jiun and others",
    title = "{New CTEQ global analysis of quantum chromodynamics with high-precision data from the LHC}",
    eprint = "1912.10053",
    archivePrefix = "arXiv",
    primaryClass = "hep-ph",
    reportNumber = "MSUHEP-19-025, PITT-PACC-1911, SMU-HEP-19-03",
    doi = "10.1103/PhysRevD.103.014013",
    journal = "Phys. Rev. D",
    volume = "103",
    number = "1",
    pages = "014013",
    year = "2021"
}

@article{Skands:2014pea,
    author = "Skands, Peter and Carrazza, Stefano and Rojo, Juan",
    title = "{Tuning PYTHIA 8.1: the Monash 2013 Tune}",
    eprint = "1404.5630",
    archivePrefix = "arXiv",
    primaryClass = "hep-ph",
    reportNumber = "CERN-PH-TH-2014-069, MCNET-14-08, OUTP-14-05P",
    doi = "10.1140/epjc/s10052-014-3024-y",
    journal = "Eur. Phys. J. C",
    volume = "74",
    number = "8",
    pages = "3024",
    year = "2014"
}

@inproceedings{Buckley:2014ctn,
    author = "Buckley, Andrew",
    title = "{ATLAS Pythia 8 tunes to 7 TeV data}",
    booktitle = "{6th International Workshop on Multiple Partonic Interactions at the LHC}",
    reportNumber = "ATL-PHYS-PROC-2014-273",
    pages = "29",
    month = "12",
    year = "2014"
}

@techreport{Hedayat1980StudyOptimality,
  author       = {Hedayat, A.},
  title        = {Study of Optimality Criteria in Design of Experiments},
  institution  = {Department of Mathematics, University of Illinois at Chicago},
  number       = {AFOSR-TR-80-0514},
  year         = {1980},
  month        = jun,
  note         = {Invited paper presented at the International Symposium on Statistics and Related Topics, Carleton University, Ottawa, Canada, May 5--8, 1980},
}

@article{Shah1960Optimality,
  author  = {Shah, K. R.},
  title   = {Optimality criteria for incomplete block designs},
  journal = {Annals of Mathematical Statistics},
  volume  = {31},
  number  = {3},
  pages   = {791--794},
  year    = {1960}
}

@article{Bierlich:2022pfr,
    author = "Bierlich, Christian and others",
    title = "{A comprehensive guide to the physics and usage of PYTHIA 8.3}",
    eprint = "2203.11601",
    archivePrefix = "arXiv",
    primaryClass = "hep-ph",
    reportNumber = "LU-TP 22-16, MCNET-22-04, FERMILAB-PUB-22-227-SCD",
    doi = "10.21468/SciPostPhysCodeb.8",
    journal = "SciPost Phys. Codeb.",
    volume = "2022",
    pages = "8",
    year = "2022"
}

@article{Chaloner:1995,
  author = {Chaloner, Kathryn and Verdinelli, Isabella},
  year = {1995},
  title = {Bayesian Experimental Design: A Review},
  journal = {Statist. Sci.},
  volume = {10},
  number = {3},
  pages = {273--304},
  publisher = {Institute of Mathematical Statistics},
  doi = {10.1214/ss/1177009939}
}

@article{Huan2024OED,  
author  = {Huan, Xun and Jagalur, Jayanth and Marzouk, Youssef},  
title   = {Optimal Experimental Design: Formulations and Computations},  
journal = {Acta Numerica},  
volume  = {33},  
year    = {2024},  
month   = sep,  
pages   = {715--840},  
doi     = {10.1017/S0962492924000023},  
url     = {https://www.osti.gov/biblio/2440515},  
note    = {OSTI ID: 2440515}
}

@article{Ilten:2022jfm,
    author = "Ilten, Phil and Menzo, Tony and Youssef, Ahmed and Zupan, Jure",
    title = "{Modeling hadronization using machine learning}",
    eprint = "2203.04983",
    archivePrefix = "arXiv",
    primaryClass = "hep-ph",
    doi = "10.21468/SciPostPhys.14.3.027",
    journal = "SciPost Phys.",
    volume = "14",
    number = "3",
    pages = "027",
    year = "2023"
}

@article{Sklar1959,
  title={Fonctions de répartition à n dimensions et leurs marges},
  author={Sklar, Abe},
  journal={Publications de l'Institut de Statistique de l'Université de Paris},
  volume={8},
  pages={229--231},
  year={1959}
}

@article{Bierlich:2023fmh,
    author = "Bierlich, Christian and Ilten, Philip and Menzo, Tony and Mrenna, Stephen and Szewc, Manuel and Wilkinson, Michael K. and Youssef, Ahmed and Zupan, Jure",
    title = "{Reweighting Monte Carlo predictions and automated fragmentation variations in Pythia 8}",
    eprint = "2308.13459",
    archivePrefix = "arXiv",
    primaryClass = "hep-ph",
    reportNumber = "FERMILAB-PUB-23-414-CSAID",
    doi = "10.21468/SciPostPhys.16.5.134",
    journal = "SciPost Phys.",
    volume = "16",
    number = "5",
    pages = "134",
    year = "2024"
}

@article{Butter:2025wxn,
    author = "Butter, Anja and others",
    title = "{Iterative HOMER with uncertainties}",
    eprint = "2509.03592",
    archivePrefix = "arXiv",
    primaryClass = "hep-ph",
    reportNumber = "FERMILAB-PUB-25-0579-CSAID",
    month = "9",
    year = "2025"
}

@article{Assi:2025avy,
    author = "Assi, Beno{\^\i}t and Bierlich, Christian and Ilten, Phil and Menzo, Tony and Mrenna, Stephen and Szewc, Manuel and Wilkinson, Michael K. and Youssef, Ahmed and Zupan, Jure",
    title = "{Characterizing the hadronization of parton showers using the HOMER method}",
    eprint = "2503.05667",
    archivePrefix = "arXiv",
    primaryClass = "hep-ph",
    reportNumber = "FERMILAB-PUB-25-0133-CSAID",
    doi = "10.21468/SciPostPhys.19.5.125",
    journal = "SciPost Phys.",
    volume = "19",
    pages = "125",
    year = "2025"
}

@article{Heller:2024onk,
    author = "Heller, Nick and Ilten, Phil and Menzo, Tony and Mrenna, Stephen and Nachman, Benjamin and Siodmok, Andrzej and Szewc, Manuel and Youssef, Ahmed",
    title = "{Rejection Sampling with Autodifferentiation - Case study: Fitting a Hadronization Model}",
    eprint = "2411.02194",
    archivePrefix = "arXiv",
    primaryClass = "hep-ph",
    reportNumber = "FERMILAB-PUB-24-0784-CSAID, MCNET-24-18",
    month = "11",
    year = "2024"
}

@article{Bierlich:2024xzg,
    author = "Bierlich, Christian and Ilten, Phil and Menzo, Tony and Mrenna, Stephen and Szewc, Manuel and Wilkinson, Michael K. and Youssef, Ahmed and Zupan, Jure",
    title = "{Describing hadronization via histories and observables for Monte-Carlo event reweighting}",
    eprint = "2410.06342",
    archivePrefix = "arXiv",
    primaryClass = "hep-ph",
    reportNumber = "FERMILAB-PUB-23-414-CSAID",
    doi = "10.21468/SciPostPhys.18.2.054",
    journal = "SciPost Phys.",
    volume = "18",
    number = "2",
    pages = "054",
    year = "2025"
}

@article{Wilkinson:2024jio,
    author = "Wilkinson, Michael K.",
    collaboration = "MLhad",
    title = "{Simulating Hadronization with Machine Learning}",
    doi = "10.1051/epjconf/202429509026",
    journal = "EPJ Web Conf.",
    volume = "295",
    pages = "09026",
    year = "2024"
}

@article{Chan:2023icm,
    author = "Chan, Jay and Ju, Xiangyang and Kania, Adam and Nachman, Benjamin and Sangli, Vishnu and Siodmok, Andrzej",
    title = "{Integrating particle flavor into deep learning models for hadronization}",
    eprint = "2312.08453",
    archivePrefix = "arXiv",
    primaryClass = "hep-ph",
    doi = "10.1103/hgbg-k7js",
    journal = "Phys. Rev. D",
    volume = "111",
    number = "11",
    pages = "116015",
    year = "2025"
}

@article{Bierlich:2023zzd,
    author = "Bierlich, Christian and Ilten, Phil and Menzo, Tony and Mrenna, Stephen and Szewc, Manuel and Wilkinson, Michael K. and Youssef, Ahmed and Zupan, Jure",
    title = "{Towards a data-driven model of hadronization using normalizing flows}",
    eprint = "2311.09296",
    archivePrefix = "arXiv",
    primaryClass = "hep-ph",
    reportNumber = "FERMILAB-PUB-23-698-CSAID",
    doi = "10.21468/SciPostPhys.17.2.045",
    journal = "SciPost Phys.",
    volume = "17",
    number = "2",
    pages = "045",
    year = "2024"
}

@article{Chan:2023ume,
    author = "Chan, Jay and Ju, Xiangyang and Kania, Adam and Nachman, Benjamin and Sangli, Vishnu and Siodmok, Andrzej",
    title = "{Fitting a deep generative hadronization model}",
    eprint = "2305.17169",
    archivePrefix = "arXiv",
    primaryClass = "hep-ph",
    doi = "10.1007/JHEP09(2023)084",
    journal = "JHEP",
    volume = "09",
    pages = "084",
    year = "2023"
}

@article{Ghosh:2022zdz,
    author = "Ghosh, Aishik and Ju, Xiangyang and Nachman, Benjamin and Siodmok, Andrzej",
    title = "{Towards a deep learning model for hadronization}",
    eprint = "2203.12660",
    archivePrefix = "arXiv",
    primaryClass = "hep-ph",
    doi = "10.1103/PhysRevD.106.096020",
    journal = "Phys. Rev. D",
    volume = "106",
    number = "9",
    pages = "096020",
    year = "2022"
}

@book{Andersson:1997xwk,
    author = "Andersson, Bo",
    title = "{The Lund Model}",
    doi = "10.1017/9781009401296",
    isbn = "978-1-009-40129-6, 978-1-009-40125-8, 978-1-009-40128-9, 978-0-521-01734-3, 978-0-521-42094-5, 978-0-511-88149-7",
    publisher = "Cambridge University Press",
    volume = "7",
    year = "1998"
}

@article{ATLAS:2024dxp,
    author = "Hayrapetyan, Aram and others",
    collaboration = "ATLAS, CMS",
    title = "{Combination of Measurements of the Top Quark Mass from Data Collected by the ATLAS and CMS Experiments at s=7 and 8~TeV}",
    eprint = "2402.08713",
    archivePrefix = "arXiv",
    primaryClass = "hep-ex",
    reportNumber = "CMS-TOP-22-001, ATLAS-TOPQ-2019-13, CERN-EP-2024-020",
    doi = "10.1103/PhysRevLett.132.261902",
    journal = "Phys. Rev. Lett.",
    volume = "132",
    number = "26",
    pages = "261902",
    year = "2024"
}

@article{Huston:2023ofk,
    author = "Huston, Joey and Rabbertz, Klaus and Zanderighi, Giulia",
    title = "{Quantum Chromodynamics}",
    eprint = "2312.14015",
    archivePrefix = "arXiv",
    primaryClass = "hep-ph",
    month = "12",
    year = "2023"
}

@article{deBlas:2025gyz,
    author = "de Blas, Jorge and others",
    title = "{Physics Briefing Book: Input for the 2026 update of the European Strategy for Particle Physics}",
    eprint = "2511.03883",
    archivePrefix = "arXiv",
    primaryClass = "hep-ex",
    reportNumber = "CERN--2025-008, CERN-ESU-2025-001",
    doi = "10.17181/CERN.35CH.2O2P",
    month = "11",
    year = "2025"
}

@article{Assi:2025gog,
    author = "Assi, Beno{\^\i}t and Bierlich, Christian and Ilten, Philip and Menzo, Tony and Mrenna, Stephen and Szewc, Manuel and Wilkinson, Michael K. and Youssef, Ahmed and Zupan, Jure",
    collaboration = "MLhad",
    title = "{Post-hoc reweighting of hadron production in the Lund string model}",
    eprint = "2505.00142",
    archivePrefix = "arXiv",
    primaryClass = "hep-ph",
    reportNumber = "FERMILAB-PUB-25-0267-CSAID, MCNET-25-07",
    doi = "10.21468/SciPostPhys.19.4.104",
    journal = "SciPost Phys.",
    volume = "19",
    number = "4",
    pages = "104",
    year = "2025"
}

@article{Thaler:2010tr,
    author = "Thaler, Jesse and Van Tilburg, Ken",
    title = "{Identifying Boosted Objects with N-subjettiness}",
    eprint = "1011.2268",
    archivePrefix = "arXiv",
    primaryClass = "hep-ph",
    reportNumber = "MIT-CTP-4191",
    doi = "10.1007/JHEP03(2011)015",
    journal = "JHEP",
    volume = "03",
    pages = "015",
    year = "2011"
}

@article{Donoghue:1979vi,
    author = "Donoghue, John F. and Low, F. E. and Pi, So-Young",
    title = "{Tensor Analysis of Hadronic Jets in Quantum Chromodynamics}",
    reportNumber = "MIT-CTP-771",
    doi = "10.1103/PhysRevD.20.2759",
    journal = "Phys. Rev. D",
    volume = "20",
    pages = "2759",
    year = "1979"
}

@article{Brandt:1964sa,
    author = "Brandt, S. and Peyrou, C. and Sosnowski, R. and Wroblewski, A.",
    title = "{The Principal axis of jets. An Attempt to analyze high-energy collisions as two-body processes}",
    doi = "10.1016/0031-9163(64)91176-X",
    journal = "Phys. Lett.",
    volume = "12",
    pages = "57--61",
    year = "1964"
}

@article{Dokshitzer:1998kz,
    author = "Dokshitzer, Yuri L. and Lucenti, A. and Marchesini, G. and Salam, G. P.",
    title = "{On the QCD analysis of jet broadening}",
    eprint = "hep-ph/9801324",
    archivePrefix = "arXiv",
    reportNumber = "IFUM-602-FT",
    doi = "10.1088/1126-6708/1998/01/011",
    journal = "JHEP",
    volume = "01",
    pages = "011",
    year = "1998"
}

@article{Catani:1992jc,
    author = "Catani, S. and Turnock, G. and Webber, B. R.",
    title = "{Jet broadening measures in $e^{+} e^{-}$ annihilation}",
    reportNumber = "CERN-TH-6570-92",
    doi = "10.1016/0370-2693(92)91565-Q",
    journal = "Phys. Lett. B",
    volume = "295",
    pages = "269--276",
    year = "1992"
}

@article{Parisi:1978eg,
    author = "Parisi, G.",
    title = "{Super Inclusive Cross-Sections}",
    reportNumber = "LPTENS 78/5",
    doi = "10.1016/0370-2693(78)90061-8",
    journal = "Phys. Lett. B",
    volume = "74",
    pages = "65--67",
    year = "1978"
}

@article{Farhi:1977sg,
    author = "Farhi, Edward",
    title = "{A QCD Test for Jets}",
    reportNumber = "HUTP-77-A059",
    doi = "10.1103/PhysRevLett.39.1587",
    journal = "Phys. Rev. Lett.",
    volume = "39",
    pages = "1587--1588",
    year = "1977"
}

@article{Field:1982dg,
    author = "Field, Richard D. and Wolfram, Stephen",
    title = "{A QCD Model for e+ e- Annihilation}",
    reportNumber = "UFTP-82-12, CALT-68-909",
    doi = "10.1016/0550-3213(83)90175-X",
    journal = "Nucl. Phys. B",
    volume = "213",
    pages = "65--84",
    year = "1983"
}

@article{Gottschalk:1983fm,
    author = "Gottschalk, Thomas D.",
    title = "{An Improved Description of Hadronization in the QCD Cluster Model for $e^+ e^-$ Annihilation}",
    reportNumber = "CALT-68-1052",
    doi = "10.1016/0550-3213(84)90253-0",
    journal = "Nucl. Phys. B",
    volume = "239",
    pages = "349--381",
    year = "1984"
}

@article{Webber:1983if,
    author = "Webber, B.R.",
    title = "{A QCD Model for Jet Fragmentation Including Soft Gluon Interference}",
    reportNumber = "CERN-TH-3713",
    doi = "10.1016/0550-3213(84)90333-X",
    journal = "Nucl. Phys. B",
    volume = "238",
    pages = "492--528",
    year = "1984"
}

@article{Andersson:1983ia,
    author = "Andersson, Bo and Gustafson, G. and Ingelman, G. and Sjostrand, T.",
    title = "{Parton Fragmentation and String Dynamics}",
    reportNumber = "LU-TP-83-10",
    doi = "10.1016/0370-1573(83)90080-7",
    journal = "Phys. Rept.",
    volume = "97",
    pages = "31--145",
    year = "1983"
}

@article{Andersson:1998tv,
    author         = "Andersson, Bo",
    title          = "{The Lund model}",
    journal        = "Camb. Monogr. Part. Phys. Nucl. Phys. Cosmol.",
    volume         = "7",
    year           = "1997",
    pages          = "1-471",
    SLACcitation   = "%%CITATION = CMPCE,7,1;%%"
}

@article{Kluth:2006bw,
    author = "Kluth, Stefan",
    title = "{Tests of Quantum Chromo Dynamics at e+ e- Colliders}",
    eprint = "hep-ex/0603011",
    archivePrefix = "arXiv",
    reportNumber = "MPP-2006-19",
    doi = "10.1088/0034-4885/69/6/R04",
    journal = "Rept. Prog. Phys.",
    volume = "69",
    pages = "1771--1846",
    year = "2006"
}

@article{Bethke:2006ac,
    author = "Bethke, Siegfried",
    title = "{Experimental tests of asymptotic freedom}",
    eprint = "hep-ex/0606035",
    archivePrefix = "arXiv",
    reportNumber = "MPE-2006-54",
    doi = "10.1016/j.ppnp.2006.06.001",
    journal = "Prog. Part. Nucl. Phys.",
    volume = "58",
    pages = "351--386",
    year = "2007"
}

@article{Benitez:2024nav,
    author = "Benitez, Miguel A. and Hoang, Andre H. and Mateu, Vicent and Stewart, Iain W. and Vita, Gherardo",
    title = "{On determining {\ensuremath{\alpha}}$_{s}$(m$_{Z}$) from dijets in e$^{+}$e$^{-}$ thrust}",
    eprint = "2412.15164",
    archivePrefix = "arXiv",
    primaryClass = "hep-ph",
    reportNumber = "MIT-CTP 5746, CERN-TH-2024-142, UWThPh2024-8",
    doi = "10.1007/JHEP07(2025)249",
    journal = "JHEP",
    volume = "07",
    pages = "249",
    year = "2025"
}

@article{dEnterria:2022hzv,
    author = "d'Enterria, D. and others",
    title = "{The strong coupling constant: state of the art and the decade ahead}",
    eprint = "2203.08271",
    archivePrefix = "arXiv",
    primaryClass = "hep-ph",
    reportNumber = "FERMILAB-CONF-22-148-T",
    doi = "10.1088/1361-6471/ad1a78",
    journal = "J. Phys. G",
    volume = "51",
    number = "9",
    pages = "090501",
    year = "2024"
}

@article{OPAL:2011aa,
    author = "Abbiendi, G. and others",
    collaboration = "OPAL",
    title = "{Determination of $alpha_s$ using OPAL hadronic event shapes at $\sqrt{s}=91$ - 209 GeV and resummed NNLO calculations}",
    eprint = "1101.1470",
    archivePrefix = "arXiv",
    primaryClass = "hep-ex",
    reportNumber = "CERN-PH-EP-2010-089",
    doi = "10.1140/epjc/s10052-011-1733-z",
    journal = "Eur. Phys. J. C",
    volume = "71",
    pages = "1733",
    year = "2011"
}

@article{ALEPH:2003obs,
    author = "Heister, A. and others",
    collaboration = "ALEPH",
    title = "{Studies of QCD at e+ e- centre-of-mass energies between 91-GeV and 209-GeV}",
    reportNumber = "CERN-EP-2003-084",
    doi = "10.1140/epjc/s2004-01891-4",
    journal = "Eur. Phys. J. C",
    volume = "35",
    pages = "457--486",
    year = "2004"
}

@article{Abbate:2010xh,
    author = "Abbate, Riccardo and Fickinger, Michael and Hoang, Andre H. and Mateu, Vicent and Stewart, Iain W.",
    title = "{Thrust at $N^{3}LL$ with Power Corrections and a Precision Global Fit for $\alpha_{s}(mZ)$}",
    eprint = "1006.3080",
    archivePrefix = "arXiv",
    primaryClass = "hep-ph",
    reportNumber = "MIT-CTP-4101, MPP-2010-7",
    doi = "10.1103/PhysRevD.83.074021",
    journal = "Phys. Rev. D",
    volume = "83",
    pages = "074021",
    year = "2011"
}

@article{Hoang:2015hka,
    author = "Hoang, Andr{\'e} H. and Kolodrubetz, Daniel W. and Mateu, Vicent and Stewart, Iain W.",
    title = "{Precise determination of $\alpha_s$ from the $C$-parameter distribution}",
    eprint = "1501.04111",
    archivePrefix = "arXiv",
    primaryClass = "hep-ph",
    reportNumber = "UWTHPH-2015-1, MIT-CTP-4630, LPN14-128",
    doi = "10.1103/PhysRevD.91.094018",
    journal = "Phys. Rev. D",
    volume = "91",
    number = "9",
    pages = "094018",
    year = "2015"
}

@article{deFlorian:2004mp,
    author = "de Florian, Daniel and Grazzini, Massimiliano",
    title = "{The Back-to-back region in e+ e- energy-energy correlation}",
    eprint = "hep-ph/0407241",
    archivePrefix = "arXiv",
    reportNumber = "CERN-PH-TH-2004-137",
    doi = "10.1016/j.nuclphysb.2004.10.051",
    journal = "Nucl. Phys. B",
    volume = "704",
    pages = "387--403",
    year = "2005"
}

@article{AKRAWY1990159,
collaboration = "OPAL",
title = {A measurement of energy correlations and a determination of $\alpha_s(M_Z^2)$ from $e^+e^-$ annihilations at $\sqrt{s}=91$\,GeV},
journal = {Physics Letters B},
volume = {252},
number = {1},
pages = {159-169},
year = {1990},
issn = {0370-2693},
doi = {https://doi.org/10.1016/0370-2693(90)91098-V},
url = {https://www.sciencedirect.com/science/article/pii/037026939091098V},
}

@article{scikit-learn,
  title={Scikit-learn: Machine Learning in {P}ython},
  author={Pedregosa, F. and Varoquaux, G. and Gramfort, A. and Michel, V.
          and Thirion, B. and Grisel, O. and Blondel, M. and Prettenhofer, P.
          and Weiss, R. and Dubourg, V. and Vanderplas, J. and Passos, A. and
          Cournapeau, D. and Brucher, M. and Perrot, M. and Duchesnay, E.},
  journal={Journal of Machine Learning Research},
  volume={12},
  pages={2825--2830},
  year={2011}
}

@inproceedings{Chen_2016,
	doi = {10.1145/2939672.2939785},
  
	url = {https://doi.org/10.1145%2F2939672.2939785},
  
	year = 2016,
	month = {aug},
  
	publisher = {{ACM}
},
  
	author = {Tianqi Chen and Carlos Guestrin},
  
	title = {{XGBoost}},
  
	booktitle = {Proceedings of the 22nd {ACM} {SIGKDD} International Conference on Knowledge Discovery and Data Mining}
}

@article{Cacciari:2011ma,
    author = "Cacciari, Matteo and Salam, Gavin P. and Soyez, Gregory",
    title = "{FastJet User Manual}",
    eprint = "1111.6097",
    archivePrefix = "arXiv",
    primaryClass = "hep-ph",
    reportNumber = "CERN-PH-TH-2011-297",
    doi = "10.1140/epjc/s10052-012-1896-2",
    journal = "Eur. Phys. J. C",
    volume = "72",
    pages = "1896",
    year = "2012"
}

@article{Cacciari:2005hq,
    author = "Cacciari, Matteo and Salam, Gavin P.",
    title = "{Dispelling the $N^{3}$ myth for the $k_t$ jet-finder}",
    eprint = "hep-ph/0512210",
    archivePrefix = "arXiv",
    reportNumber = "LPTHE-05-32",
    doi = "10.1016/j.physletb.2006.08.037",
    journal = "Phys. Lett. B",
    volume = "641",
    pages = "57--61",
    year = "2006"
}

@article{10.1098/rsta.1933.0009,
    author = {Neyman, Jerzy and Pearson, Egon Sharpe},
    title = {{IX. O}n the problem of the most efficient tests of statistical hypotheses},
    journal = {Philosophical Transactions of the Royal Society of London, Series A: Containing Papers of a Mathematical or Physical Character},
    volume = {231},
    number = {694-706},
    pages = {289-337},
    year = {1933},
    month = {02},
    issn = {0264-3952},
    doi = {10.1098/rsta.1933.0009},
    url = {https://doi.org/10.1098/rsta.1933.0009},
    eprint = {https://royalsocietypublishing.org/rsta/article-pdf/231/694-706/289/240090/rsta.1933.0009.pdf},
}

@article{Catani:1991hj,
    author = "Catani, S. and Dokshitzer, Yuri L. and Olsson, M. and Turnock, G. and Webber, B. R.",
    title = "{New clustering algorithm for multi - jet cross-sections in e+ e- annihilation}",
    reportNumber = "CAVENDISH-HEP-91-5",
    doi = "10.1016/0370-2693(91)90196-W",
    journal = "Phys. Lett. B",
    volume = "269",
    pages = "432--438",
    year = "1991"
}

@article{Rizvi:2023mws,
    author = "Rizvi, Shahzar and Pettee, Mariel and Nachman, Benjamin",
    title = "{Learning likelihood ratios with neural network classifiers}",
    eprint = "2305.10500",
    archivePrefix = "arXiv",
    primaryClass = "hep-ph",
    doi = "10.1007/JHEP02(2024)136",
    journal = "JHEP",
    volume = "02",
    pages = "136",
    year = "2024"
}

@article{Cranmer:2015bka,
    author = "Cranmer, Kyle and Pavez, Juan and Louppe, Gilles",
    title = "{Approximating Likelihood Ratios with Calibrated Discriminative  Classifiers}",
    eprint = "1506.02169",
    archivePrefix = "arXiv",
    primaryClass = "stat.AP",
    month = "6",
    year = "2015"
}

@article{Coccaro:2019lgs,
    author = "Coccaro, Andrea and Pierini, Maurizio and Silvestrini, Luca and Torre, Riccardo",
    title = "{The DNNLikelihood: enhancing likelihood distribution with Deep Learning}",
    eprint = "1911.03305",
    archivePrefix = "arXiv",
    primaryClass = "hep-ph",
    reportNumber = "CERN-TH-2019-187",
    doi = "10.1140/epjc/s10052-020-8230-1",
    journal = "Eur. Phys. J. C",
    volume = "80",
    number = "7",
    pages = "664",
    year = "2020"
}

@article{Cranmer_sbi_review,
author = {Kyle Cranmer  and Johann Brehmer  and Gilles Louppe },
title = {The frontier of simulation-based inference},
journal = {Proceedings of the National Academy of Sciences},
volume = {117},
number = {48},
pages = {30055-30062},
year = {2020},
doi = {10.1073/pnas.1912789117},
URL = {https://www.pnas.org/doi/abs/10.1073/pnas.1912789117},
eprint = {https://www.pnas.org/doi/pdf/10.1073/pnas.1912789117},
}

@article{ATLAS:2025clx,
    author = "Aad, Georges and others",
    collaboration = "ATLAS",
    title = "{An implementation of neural simulation-based inference for parameter estimation in ATLAS}",
    eprint = "2412.01600",
    archivePrefix = "arXiv",
    primaryClass = "physics.data-an",
    reportNumber = "CERN-EP-2024-305",
    doi = "10.1088/1361-6633/add370",
    journal = "Rept. Prog. Phys.",
    volume = "88",
    number = "6",
    pages = "067801",
    year = "2025"
}

@article{Brehmer:2018eca,
    author = "Brehmer, Johann and Cranmer, Kyle and Louppe, Gilles and Pavez, Juan",
    title = "{A Guide to Constraining Effective Field Theories with Machine Learning}",
    eprint = "1805.00020",
    archivePrefix = "arXiv",
    primaryClass = "hep-ph",
    doi = "10.1103/PhysRevD.98.052004",
    journal = "Phys. Rev. D",
    volume = "98",
    number = "5",
    pages = "052004",
    year = "2018"
}

@article{Brehmer:2018kdj,
    author = "Brehmer, Johann and Cranmer, Kyle and Louppe, Gilles and Pavez, Juan",
    title = "{Constraining Effective Field Theories with Machine Learning}",
    eprint = "1805.00013",
    archivePrefix = "arXiv",
    primaryClass = "hep-ph",
    doi = "10.1103/PhysRevLett.121.111801",
    journal = "Phys. Rev. Lett.",
    volume = "121",
    number = "11",
    pages = "111801",
    year = "2018"
}

@misc{deistler2025simulationbasedinferencepracticalguide,
      title={Simulation-Based Inference: A Practical Guide}, 
      author={Michael Deistler and Jan Boelts and Peter Steinbach and Guy Moss and Thomas Moreau and Manuel Gloeckler and Pedro L. C. Rodrigues and Julia Linhart and Janne K. Lappalainen and Benjamin Kurt Miller and Pedro J. Gonçalves and Jan-Matthis Lueckmann and Cornelius Schröder and Jakob H. Macke},
      year={2025},
      eprint={2508.12939},
      archivePrefix={arXiv},
      primaryClass={stat.ML},
      url={https://arxiv.org/abs/2508.12939}, 
}

@article{Schyns:1996xrr,
    author = "Schyns, E.",
    title = "{NEWTAG -  {\textbackslash}pi , K, p Tagging for Delphi RICHes}",
    reportNumber = "DELPHI-96-103 RICH 89",
    month = "7",
    year = "1996"
}

\nolinenumbers

%%%%%%%%%%%%%%%%%%%%%%%%%%%%%%%%%%%%%%%%%
\appendix
%%%%%%%%%%%%%%%%%%%%%%%%%%%%%%%%%%%%%%%%%
\section{Observable definitions}
\label{sec:observable_definition}

In this appendix, we collect the definitions of the 15 hadronization-sensitive observables used in our numerical studies (see also \cref{eq:list:O} in the main text).

\paragraph{Multiplicities: $n_{\rm had}$, $n_{\rm ch}$, $n_{\rm baryon}$, $n_{\rm str}$.} 
The hadron multiplicity $n_{\rm had}$ counts the total number of hadrons in an event:
\beq 
n_{\rm had}=\sum_{i\in {\rm hadrons}} 1.
\eeq
Similarly, $n_{\rm ch}$ counts the total number of hadrons with nonzero electric charge (not to be confused with the total charge, which can be either positive or negative), and $n_{\rm baryon}$ counts the total number of hadrons with nonzero baryon number (similarly, not to be confused with the total baryon number):
\beq 
n_{\rm ch}=\sum_{i\in {\rm hadrons}} \Theta(|Q_i| ),  \qquad n_{\rm baryon} = \sum_{i\in {\rm hadrons}} \Theta(|B_i|  ),
\eeq
where $Q_i$ and $B_i$ are the electric charge and baryon number of particle $i$, and $\Theta$ is the usual Heaviside Theta function.
Note that for $n_{\rm ch}$, charged leptons are not included.
Finally $n_{\rm str}$ counts the absolute strangeness:
\beq 
n_{\rm str}=\sum_{i\in {\rm hadrons}} |s_i-\bar{s}_i|,
\eeq
where $s_i$ ($\bar{s}_i$) count the number of strange (antistrange) quarks in hadron $i$.

\paragraph{Event shapes: $1-T$, $B_T$, $B_W$, $C$, $D$.} 
Thrust $T$ is defined as~\cite{Brandt:1964sa,Farhi:1977sg}
\begin{equation}
  T = \max_{\vec{n}_T}\frac{\sum_i |\vec{p}_i \cdot \vec{n}_T|}{\sum_i |\vec p_i|},
\end{equation}
where the sum runs over all particles $i$ in the event with three-momenta $\vec{p}_i$, and the unit vector $\vec{n}_T$ is chosen to maximize this expression. Thrust ranges from $T = 0.5$ for isotropic events to $T = 1$ for perfectly collimated dijets.

The thrust axis $\vec{n}_T$ divides the event into two hemispheres $S_\pm$, used to define jet broadenings~\cite{Catani:1992jc,Dokshitzer:1998kz}:
\begin{equation}
  B_\pm =\frac{\sum_{i \in S_\pm}|\vec{p}_i \times \vec{n}_T|}{2\sum_i |\vec{p}_i|}.
\end{equation}
The total broadening is $B_T = B_+ + B_-$ and the wide jet broadening is $B_W = \max(B_+, B_-)$.

The $C$ and $D$ parameters are defined from eigenvalues of the linearized momentum tensor~\cite{Parisi:1978eg,Donoghue:1979vi}
\begin{equation}
  \Theta^{ab} = \frac{1}{\sum_i |\vec p_i|} \sum_i \frac{p_i^a p_i^b}{|\vec p_i|}, \qquad a,b=1,2,3,
\end{equation}
where $p_i^a$ denotes the $a$-th component of $\vec{p}_i$. If $\lambda_1 \geq \lambda_2 \geq \lambda_3$ are the eigenvalues of $\Theta^{ab}$, then
\begin{equation}
  C = 3(\lambda_1\lambda_2 + \lambda_2\lambda_3 + \lambda_3\lambda_1), \qquad D = 27\lambda_1\lambda_2\lambda_3.
\end{equation}

\paragraph{Jet observables: $n_{\rm jet}$, $\tau_1$, $\tau_2$, $\tau_3$.} 
Jets are clustered using the generalized $k_T$ algorithm~\cite{Catani:1991hj,Cacciari:2005hq,Cacciari:2011ma} implemented in \texttt{FastJet}. Specifically, we use \texttt{ee\_gen\_kt\_algorithm} with $R = 1$ and $p = -1$, requiring jet energies $E > 5$ GeV to exclude soft jets with few constituents. The number of clustered jets in an event is $n_{\rm jet}$.

The $N$-subjettiness $\tau_N$ quantifies whether a jet contains $N$ distinct subjets~\cite{Thaler:2010tr}. For each jet, we re-cluster its constituents into $N$ subjets using the $k_T$ algorithm with $R = 1$, then compute
\beq
\tau_N=\frac{1}{d_0}\sum_k p_{T,k} \min\{\Delta R_{1,k}, \Delta R_{2,k},\ldots, \Delta R_{N,k}\},
\eeq
where the sum runs over all particles $k$ in the jet, $p_{T,k}$ is the transverse momentum of particle $k$ relative to the jet axis, $\Delta R_{n,k}$ is the angular distance between particle $k$ and subjet $n$ in the rapidity-azimuth plane, and $d_0 = R\sum_k p_{T,k}$ normalizes the observable. Small $\tau_N$ indicates $N$ well-separated subjets, while large $\tau_N$ suggests more than $N$ subjets.

\paragraph{Correlators: EEC, NNC.} 
The energy-energy correlator (EEC) measures the angular distribution of energy flow~\cite{deFlorian:2004mp}:
\beq
\label{eq:EEC:def}
\text{EEC}(\theta) = \frac{1}{\Delta\theta N_{\rm ev}} \sum_{\rm events}^{N_{\rm ev}} \sum_{i<j} \frac{E_i E_j}{E_{\rm vis}^2} \, \Theta(\theta - \Delta\theta/2 < \theta_{ij} < \theta + \Delta\theta/2),
\eeq
where $N_{\rm ev}$ is the number of events, $\theta_{ij}$ is the angle between particles $i$ and $j$, $E_{\rm vis} = \sum_i E_i$ is the total visible energy, and $\Theta$ denotes the indicator function selecting particle pairs in the angular bin of width $\Delta\theta$ centered at $\theta$.

The number-number correlator (NNC) is defined analogously but counts particle pairs rather than weighting by energy:
\beq
\label{eq:NNC:def}
\text{NNC}(\theta) = \frac{1}{\Delta\theta N_{\rm ev}} \sum_{\rm events}^{N_{\rm ev}} \sum_{i<j} \frac{1}{n_{\rm had}^2} \, \Theta(\theta - \Delta\theta/2 < \theta_{ij} < \theta + \Delta\theta/2),
\eeq
where $n_{\rm had}$ is the hadron multiplicity of the event, so $1/n_{\rm had}$ represents the number fraction carried by each particle.

%%%%%%%%%%%%%%%%%%%%%%%%%%%%%%%%%%%%%%%%%%%%%%%%%%
%%%%%%%%%%%%%%%%%%%%%%%%%%%%%%%%%%%%%%%%%%%%%%%%%%
%%%%%%%%%%%%%%%%%%%%%%%%%%%%%%%%%%%%%%%%%%%%%%%%%%
\section{Observable rankings with fixed overlap penalty}
\label{app:fixed_beta}

\begin{figure}[t!]
\centering
\includegraphics[width=0.9\textwidth]{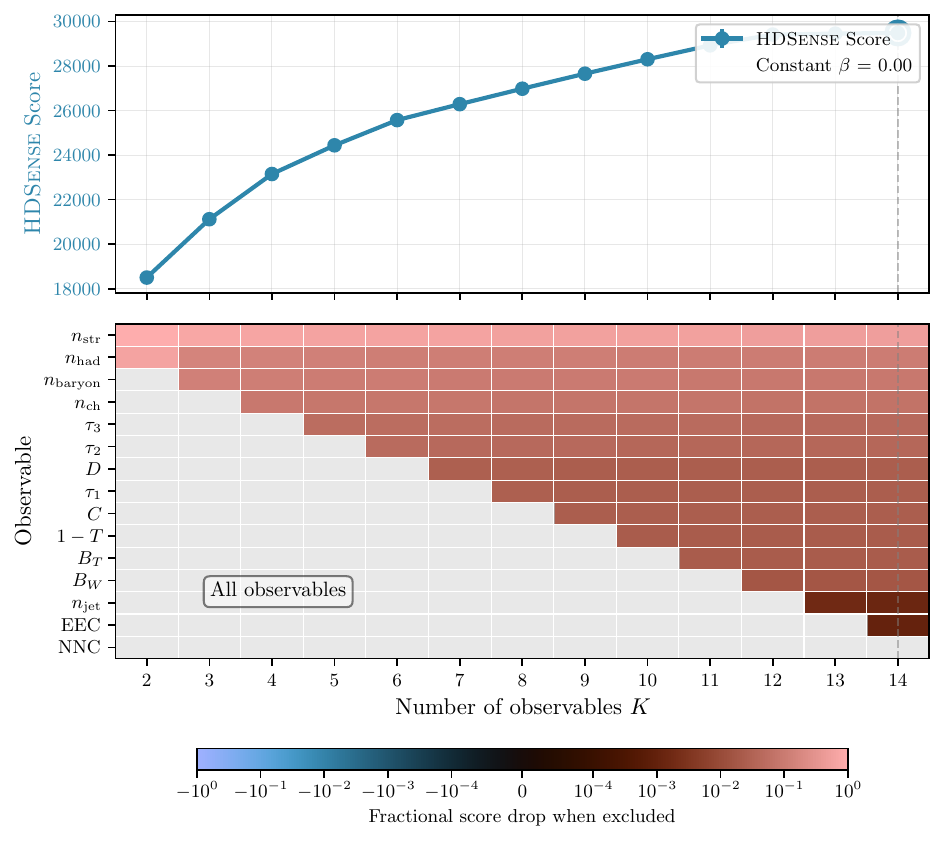}
\caption{\method ranking for all 15 observables with $\beta = 0$ (no overlap penalty). Layout as in \cref{fig:all_observables_no_detector}. The score increases monotonically, providing no indication of when additional observables cease providing meaningful constraints due to correlations with already-selected measurements.
}
\label{fig:all_obs_beta0}
\end{figure}

The main text uses an adaptive hyperparameter $\beta(K)$ that varies with the number of selected observables $K$, as defined in \cref{eq:heuristic}. This adaptive choice leads to a monotonic subset growth in the numerical examples studied in this work, \ie, the optimal $K$-observable subset was found  always to be contained within the optimal $(K+1)$-observable subset. In this appendix we show instead the results for fixed $\beta$ values to illustrate the consequences of non-adaptive penalties.

\Cref{fig:all_obs_beta0,fig:all_obs_beta05} show results for all 15 observables with  $\beta = 0$ (no overlap penalty) and $\beta = 0.5$ (moderate fixed penalty), respectively. The color coding is as in \cref{fig:all_observables_no_detector}, with the observables that are not selected in the optimal subset at each $K$ greyed out, and with darker colors of the boxes indicating more important observables. 

For $\beta = 0$, \cref{fig:all_obs_beta0}, the score monotonically increases with $K$, since ${\mathcal S}_\text{HD}$ now simply sums traces of Fisher information matrices without penalizing overlaps. This provides no guidance on when to stop adding observables: the score suggests that every additional measurement meaningfully improves constraints, even when observables are highly correlated and provide redundant information. The constant increase fails to capture the diminishing returns from correlated measurements.

For $\beta = 0.5$, \cref{fig:all_obs_beta05}, the score peaks at $K = 2$ and becomes negative from $K = 10$ onwards. If ${\mathcal S}_\text{HD}$ is interpreted as an proxy for $A$-optimality score based on full information matrix, $\Tr (I_\text{full})$, then the negative scores are unphysical. More precisely, \method approximates the trace of a profiled Fisher information matrix, $I_{\rm pr.}(\bm \theta)$ defined in \cref{eq:profiled_fisher} in the main text, which is non-negative. A negative score indicates a too large overlap penalty $2\beta\mathcal{P}_{\text{overlap}}$ that exceeds unity, driving the approximation outside its regime of validity, where it can no longer represent a Fisher information matrix. This breakdown occurs because the choice of a fixed penalty parameter $\beta$ eventually over-compensates the effect of correlations as they grow with increasing $K$. The use of adaptive $\beta(K)$ in \cref{eq:heuristic}, which was adopted in the main text, avoids this by decreasing $\beta$ as $K$ increases, maintaining validity across all subset sizes.

\begin{figure}[t!]
\centering
\includegraphics[width=0.9\textwidth]{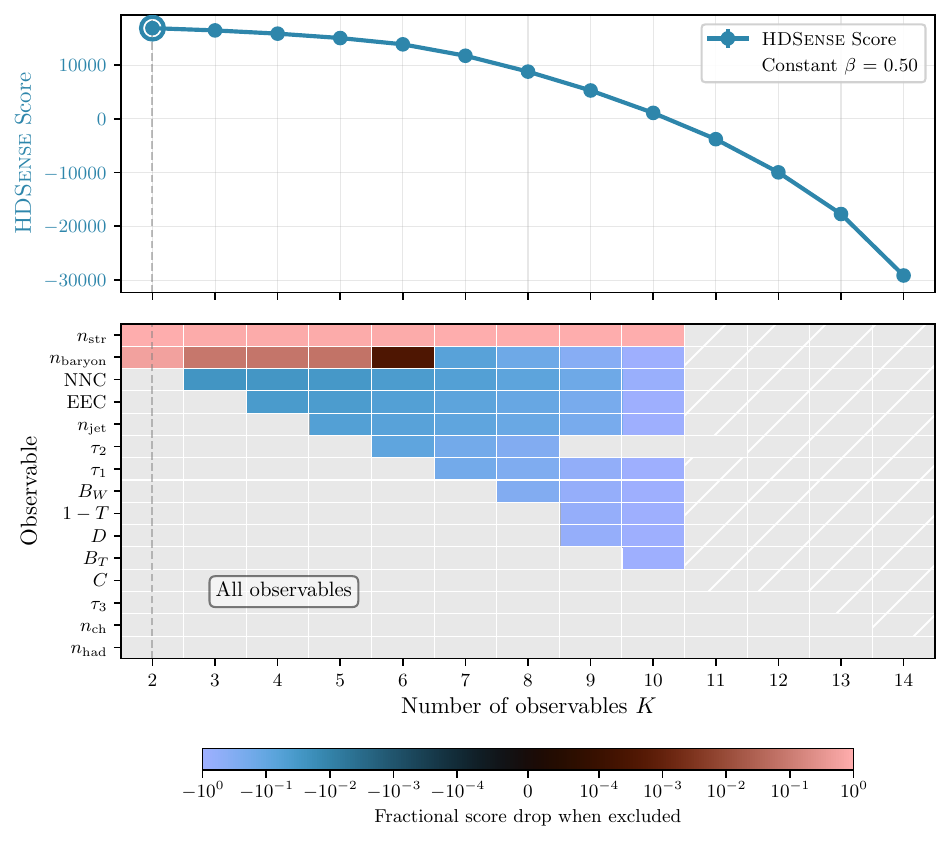}
\caption{\method ranking for all 15 observables with fixed $\beta = 0.5$. Layout as in \cref{fig:all_observables_no_detector}. The score peaks at $K = 2$, becoming negative (dark grey shading) from $K = 10$ onwards. Negative scores indicate the approximation no longer represents a valid Fisher information matrix. Selected subsets (shown with white dashed lines for $K = 10$ onwards) do not grow monotonically.
}
\label{fig:all_obs_beta05}
\end{figure}

%%%%%%%%%%%%%%%%%%%%%%%%%%%%%%%%%%%%%%%%%%%%%%%%%%
%%%%%%%%%%%%%%%%%%%%%%%%%%%%%%%%%%%%%%%%%%%%%%%%%%
%%%%%%%%%%%%%%%%%%%%%%%%%%%%%%%%%%%%%%%%%
\section{Relation to optimal experiment design scores}
\label{app:sec:determinant:score}

The \method score ${\mathcal S}_\text{HD}$, introduced in the main text, shares some attributes with, yet is also distinct from, the so called classical alphabetic optimality design criteria in the theory of optimal experiment design~\cite{Chaloner:1995,Huan2024OED}. 
Here, many different scalar criteria are defined to be optimized, corresponding to some notion of maximal information extraction in a measurement.
The two most relevant for our purposes are the $A$-optimality criterion and $D$-optimality criterion, defined as \cite{kiefer1959optimum},\footnote{``$A$'' for ``average'', or trace, and ``$D$'' for ``determinant''.}
\begin{align}
    A &= \Tr\big(I^{-1}\big), \qquad D = \det\big(I^{-1}\big),
\end{align}
where $I$ is the full Fisher information matrix.
Smaller values of $A$ or $D$ are considered more optimal, \ie, the $A$($D$)-optimality criterion seeks to minimize $A$($D$) score.
By the Cram\'er-Rao bound, $A$-optimality corresponds to finding a subset of observables that minimizes the average variance of the estimated parameters, and $D$-optimality corresponds to minimizing the volume of the error ellipsoid---or equivalently, maximizing the differential Shannon information.
A distinguishing feature of the $D$-optimality criterion is that it is invariant under the rescaling of parameters. That is, if one redefines parameters via a linear transformation $\bm \theta'= R \bm \theta$, where $R$ is some $N_\text{par}\times N_\text{par}$ matrix, the $D$ score simply gets shifted by a constant, $D'=D+2\log \det R$. The optimal subsets chosen by maximizing $D$ or by maximizing $D'$ are therefore the same. 
%\rikab{Is this not also true for $A$ score?} \jz{It is true for unitary $R$ bot not in general} \rikab{Do we need this property? }

The ${\mathcal S}_\text{HD}$ score is based on optimizing $\Tr[I]$ rather than $\Tr[I^{-1}]$, which is related to but not identical to $A$-optimality.
This can be considered a third criterion, though $\Tr[I]$-optimality is normally not associated with its own standardized letter. 
They are related via,
\begin{align}
    \Tr\big(I^{-1}\big) \geq \frac{N_{\rm par}^2}{\Tr[I]}.
\end{align}
In \cref{sec:derivation}, we showed how ${\mathcal S}_\text{HD}$ provides an approximate lower bound on our ``trace''-optimality, which in turn gives a bound on $A$-optimality. 
While they are not the same, $\Tr[I]$ and $\Tr[I^{-1}]$ are highly correlated in practice.

Note that the $A$ score and the (negative-log) $D$ score were introduced in \cref{sec:validation} as criteria to validate the ${\mathcal S}_\text{HD}$. 
The results in \cref{sec:validation} imply that ${\mathcal S}_\text{HD}$, though based on $\Tr[I]$, performs similarly to both the $A$-criterion and $D$-criterion, but is much easier to compute.

Beyond $A$-optimality and $D$-optimality, there are other alphabetic design criteria used in experimental design, which one may also can consider.
For example, 
the so called $E$-optimality criterion seeks to minimize the maximum eigenvalue of $I^{-1}$:
\begin{align}
    E = \lambda_{\max}(I^{-1}).
\end{align}
That is, $E$-optimality seeks to reduce the uncertainty on the least constrained direction in the parameter space. Even though $E$-optimality, unlike $A$-optimality, $D$-optimality, $\Tr[I]$ and $\mathcal{S}_{\rm HD}$, is based on just a single eigenvalue of Fisher information matrix, it indirectly has implications for other eigenvalues---requiring $E$-optimality 
%This then 
necessarily also means that all the other directions are reasonably constrained. 

An interesting limiting case is the so called ``symmetric design'', where the observables are chosen such that all the eigenvalues of the Fisher information matrix are equal, \ie, $\lambda_1=\lambda_2=\cdots=\lambda_{N_\text{par}}$. 
If symmetric design can be achieved for the maximal value of $A$, this choice of observables is also $D$- and $E$-optimal. 
$S$-optimality is based on this observation: for a constant $\Tr[I]$, $S$-optimality minimizes \cite{Hedayat1980StudyOptimality,Shah1960Optimality}
\beq
\Phi(I)=\sum_i \lambda_i^2.
\eeq
For fixed $\Tr[I]$, this is equivalent to minimizing the Euclidean distance between the vector of eigenvalues of the Fisher information matrix, $(\lambda_1, \ldots, \lambda_{N_\text{pr}})$, and the would-be vector of eigenvalues of $I$ that one would get in the symmetric design case, $(\Tr[I], \ldots, \Tr[I])/N_\text{par}$.
That is, $S$-optimality corresponds to finding an observable configuration that attempts to simultaneously optimize the $A$-, $D$-, and $E$-criteria, since this only occurs at the symmetric design point.
A related optimality condition is the so-called $(M,S)$-optimality, where among the choices for the sets of observables that maximize $\Tr[I]$, one chooses the one that minimizes $\Phi(I)$, which corresponds to ``greedily'' optimizing the trace criterion before then optimizing $D$ and $E$.

With the many choices for the optimal design criteria that use the full Fisher information matrix, it is clear that one could also entertain alternative definition of the \method score. For instance, instead of the
$\mathrm{Info}(\mathcal{X})$ prefactor in ${\mathcal S}_\text{HD}$, \cf~\cref{eq:bounded-score},  which was based on the trace of the Fisher information matrices, one could have equally well used the determinants instead. 
However, we note that the determinant has several disadvantages: it vanishes for singular matrices (when observables fail to constrain all parameters), making it ill-defined without regularization. 
Even with regularization, the determinant is extremely sensitive to the smallest eigenvalue of the Fisher matrix—a single poorly-constrained parameter direction causes the determinant to collapse regardless of how well other directions are determined. This makes determinant-based scores numerically unstable and difficult to interpret. 
The trace, measuring average constraint strength across all parameter directions, is more robust and degrades gracefully when some parameters are weakly constrained.
Finally, determinants are generically more computationally expensive than traces, as they scale as $\mathcal{O}(n^{2-3})$, which can be relevant as we consider a combinatorially large number of observable sets. Given the greater robustness, interpretability, and computational stability of the trace, we adopt it as our primary score.

\end{document}